\documentclass[AMA,STIX1COL,natbib]{WileyNJD-v2}
\usepackage{moreverb}
\usepackage{NJDnatbib}
\usepackage{subfigure}

\numberwithin{equation}{section}

\newcommand\BibTeX{{\rmfamily B\kern-.05em \textsc{i\kern-.025em b}\kern-.08em
T\kern-.1667em\lower.7ex\hbox{E}\kern-.125emX}}

\articletype{PREPRINT}%

\received{}
\revised{}
\accepted{}
\copyyear{2025}

\begin{document}

\title{Estimating the Number of Opioid Overdoses in British Columbia Using Relational Evidence with Tree Structure \protect}

\author[1,2]{Mallory J Flynn}

\author[1]{Paul Gustafson}

\author[2]{Michael A. Irvine}

\authormark{Mallory J Flynn \textsc{et al}}

\address[1]{\orgdiv{Department of Statistics}, \orgname{University of British Columbia}, \orgaddress{\state{Vancouver, BC}, \country{Canada}}}

\address[2]{\orgname{British Columbia Centre for Disease Control}, \orgaddress{\state{Vancouver, BC}, \country{Canada}}}

\corres{Mallory J Flynn, \email{mallory.flynn@stat.ubc.ca}}

\presentaddress{Department of Statistics, Faculty of Science, University of British Columbia, 2207 Main Mall, Vancouver, BC, V6T 1Z4, Canada}

\abstract[Abstract]{In many fields, populations of interest are hidden from data for a variety of reasons, though their magnitude remains important in determining resource allocation and appropriate policy. In public health and epidemiology, linkages or relationships between sources of data may exist due to intake structure of care providers, referrals, or other related health programming. These relationships often admit a tree structure, with the target population represented by the root, and paths from root-to-leaf representing pathways of care after a health event.  
In the Canadian province of British Columbia (BC), significant efforts have been made in creating an opioid overdose cohort, a tree-like linked data structure which tracks the movement of individuals along pathways of care after an overdose. In this application, the root node represents the target population, the total number of overdose events occurring in BC during the specified time period.  We compare and contrast two methods of estimating the target population size - a weighted multiplier method based on back-calculating estimates from a number of paths and combining these estimates via a variance-minimizing weighted mean, and a fully Bayesian hierarchical model.}

\keywords{Bayesian modeling; multiplier method; population estimation; tree; overdose; opioid}

\jnlcitation{\cname{%
\author{M.J. Flynn}, and
\author{P. Gustafson}, and \author{M.A. Irvine}} (\cyear{2025}), 
\ctitle{Estimating the Number of Opioid Overdoses in British Columbia Using Relational Evidence with Tree Structure}, \cjournal{Preprint}, \cvol{2025;01:1--21}.}

\maketitle

\footnotetext{\textbf{Abbreviations:} MM, multiplier method; WMM, weighted multiplier method; ATD, acute toxicity death}

\section{Introduction}\label{intro}

The drug toxicity public health emergency continues to be one of the most significant public health challenges of the 21st century, with more than 44,000 Canadians dying from opioid overdose between January 2016 and December 2023, with rapid increases in drug overdoses and deaths throughout North America \cite{fischeropioid2014,guyopioid2017,janjuaopioid2018, overdoses2022}.  While in many jurisdictions increased numbers of drug toxicity deaths were driven by prescription opioids \cite{prescriptionopioidUS}, in British Columbia a major contributor to the increasing rate of drug toxicity deaths was the introduction of fentanyl and fentanyl analogues to the illicit drug supply \cite{baldwinfent}.  During the COVID-19 pandemic, opioid-related overdoses again increased sharply throughout, with an average of 22 overdose deaths per day observed between January and December 2023, compared with eight deaths per day in 2016 and 12 deaths per day, nationally, in 2018 \cite{overdoses2022}.  Approximately 89\% of these deaths occurred in BC, Alberta, and Ontario alone, and a similar percentage of accidental, apparent overdose toxicity deaths are occurred among individuals between the ages of 20 and 59 years \cite{overdoses2022} - significantly lower than the 82 year average life expectancy of Canadians \cite{lifeexp2020}. 

In British Columbia, there has been a large increase in the number of deaths due to opioid-related overdose \cite{bccoronerfentanyl2017,bccoronerdrug2017} and a corresponding yearly drop in life expectancy between 2014 and 2018 attributed to these deaths \cite{statcan2018,yeopioid2018}.  Although research estimates suggest prevention methods have aided in averting deaths due to opioid toxicity \cite{irvinethnkits2018,irvinevarbayes2019}, the total numbers of overdose events remain unclear, and in British Columbia, the number of attributed to opiates, in particular, is conjectured to be dramatically under-represented by healthcare data alone \cite{lauraopioiddata}.  
Overdose events are not fully captured in health administrative data sources, due in part to bystander administration of naloxone, persons not seeking medical care, and miscoded data.  As such, the true number of overdoses is under-reported, and it is probable that there is a much higher burden of overdose when non-fatal overdose events are considered alongside mortality data \cite{opioidcanada2019,macdougallcohort2019}.  Synthesis of available data sources, including existing surveillance data and  estimates from the literature, is required to produce indirect statistical estimates of the total number of overdoses \cite{lauraopioiddata}.  A comprehensive and scalable method to estimate the overall number of opioid overdose events in the population is instrumental in improving the health of communities by providing important information to decision makers and clinicians to inform policies and resource allocation, programs, intervention strategies, and clinical care.

\begin{figure}
	\centering
	\includegraphics[width=.9\linewidth]{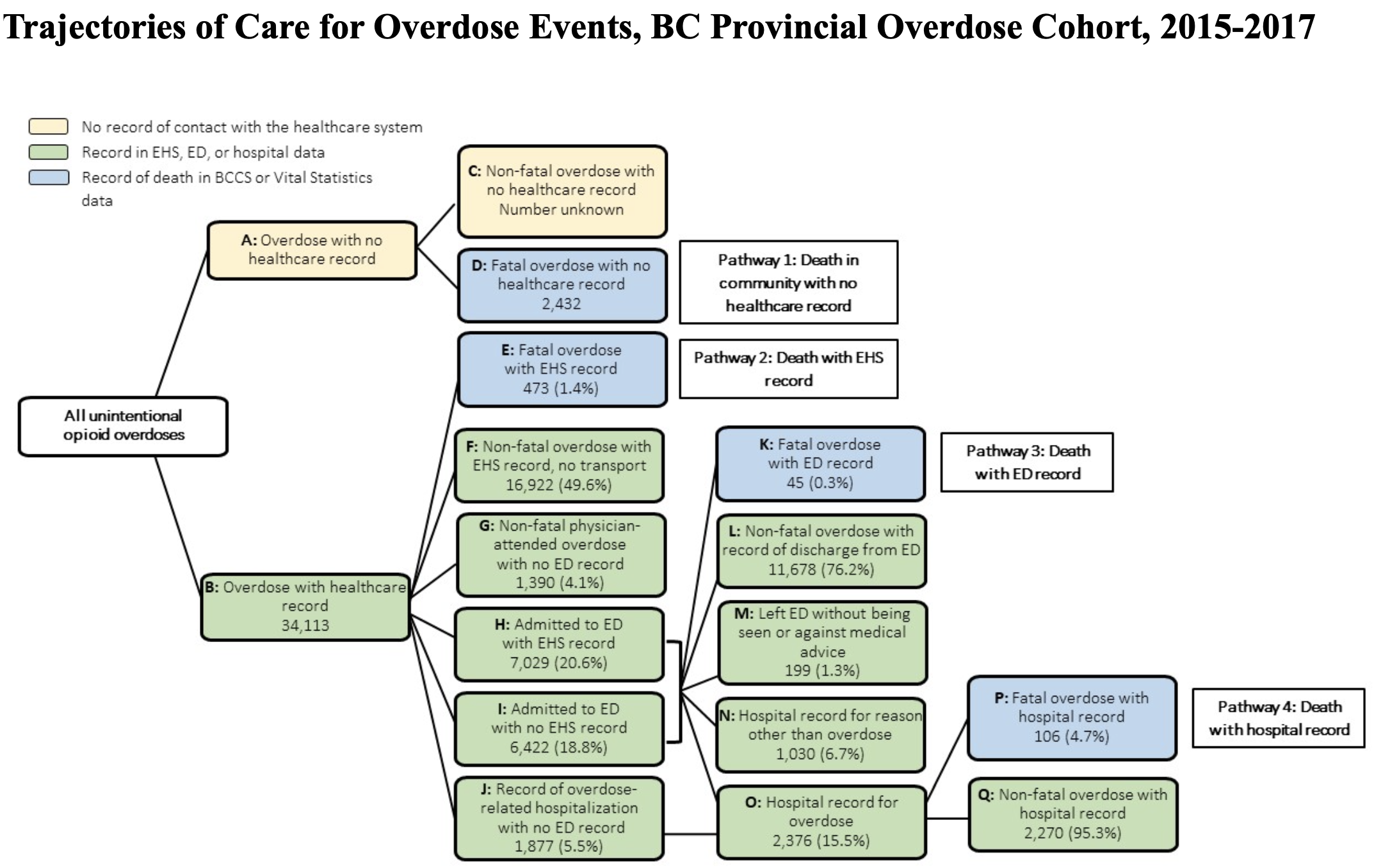}
	\caption{Opioid overdose reporting pathways tree.  Created by PHAC and BCCDC, Nov. 2021 \cite{opioiddata}.}
	\label{fig:pathwaystree}
\end{figure}

The Public Health Agency of Canada (PHAC) and the BC Centre for Disease Control (BCCDC) have created a dataset that is very well suited for addressing this issue \cite{macdougallcohort2019}, which links individuals existing in multiple health datasets (including ambulance data, emergency room data, hospital admissions, coroners data, and others) and tracks their progression through a tree of possible treatment pathways after an overdose event (see Figure \ref{fig:pathwaystree}).  It is important to note that while all of the nodes in Figure \ref{fig:pathwaystree} emanating from the lower child of the root node contain counts, many of these counts are simply summations of the observed counts of descendents (e.g., node $B$ is a summation of leaf descendents), with parent nodes themselves unobserved.  In addition, measured nodes are only partially observed; estimates of the proportion of individuals missing have been determined through preliminary investigations of the missingness expected in these sources of data.

The traditional multiplier method is a popular means of target population size estimation in public health and epidemiology, as it is applicable to existing data such as health administrative data.  When knowledge about multiple sub-populations of the target population is available, the weighted multiplier method (WMM) is an extension of the traditional methodology that allows for multiple sources of evidence to be utilized and synthesized into a single estimate of the target population, which is represented in the root of the tree that is constructed by the combined evidence \cite{flynnmethods}. The pathways of care elucidated by the Provincial Overdose Cohort (POC) data regarding the overdose crisis in British Columbia presents an ideal application for the WMM due to the clear creation of a tree structure with multiple known sub-populations which are mutually exclusive.  It represents pathways of care after an overdose event; while one major arm of the tree is well-informed using healthcare administrative datasets, the other major arm is largely unobserved, as these nodes represent those individuals that did not access healthcare services after an overdose event. Here, we apply both the WMM and a Bayesian hierarchical model to this data structure in order to estimate the total number of overdose events over the given time period, and compare and contrast the estimates provided by each method. In a companion paper, we describe the underlying theory behind both the multiplier-based methodology and the hierarchical Bayesian model \cite{flynnmethods}. In another companion paper, we describe software packages developed to ease implementation of the novel multiplier-based methodology, as well as a hierarchical Bayesian model adaptable to any general tree structure \cite{flynncomp}.

The POC data are owned by the BC Provincial Health Authorities and the Ministry of Health and stewarded by the BCCDC \cite{bccohortdata, pnetdata, nacrsdata, daddata, mspdata}.  Access to data provided by the Data Steward(s) is subject to approval, but can be requested for research projects through the Data Steward(s) or their designated service providers. All inferences, opinions, and conclusions drawn in this thesis are those of the author(s), and do not reflect the opinions or policies of the Data Steward(s).

\section{Methods} 
A wealth of literature exists demonstrating the utility of the multiplier method and Bayesian modeling to provide accurate population size estimates and corresponding uncertainty.  The multiplier method, in particular, is commonly used by public health agencies and institutes globally in the estimation of the size of key populations such as populations at higher risk of blood-borne infectious disease transmission (e.g., HIV, hepatitis), including people who currently or formerly inject drugs \cite{antaIDU2010,deangelisIDU2004,khalidIDU2014} or men who have sex with men \cite{birrellHIV2013,richMSM2017,pazbailey2011,khalidIDU2014,johnstonHIV2011}.  Similarly, Bayesian modeling is widely used to inform unknown population sizes in these same applications \cite{heesterbeekreview2015,prevostHCV2015,sweetingIDU2009} and is also used to estimate the impact of intervention strategies or therapies \cite{irvinevarbayes2019,irvinethnkits2018,irvineHIV2018}. In the following, we analyze the POC data to obtain an estimate of the total number of opioid overdose events in British Columbia during the 2015 to 2017 time period.  Two methods of estimation are used; a hierarchical Bayesian model is created and implemented in JAGS, and compared with an estimate obtained using the WMM \cite{flynnmethods}.  Implementation is achieved using the \texttt{R} packages \textit{AutoWMM} and \textit{JAGStree} \cite{flynncomp}.

\subsection{Hypotheses}
Due to the nature of the data, which is based on a single set of linked observations from a variety of evidence sources, we expect a Bayesian model to rely heavily on the prior inputs.  A sensitivity analysis is undertaken to assess the extent of the effects of priors.  We expect a WMM approach on the data to provide similar estimates to the Bayesian modeling, though the total number of overdose events may be underestimated by the WMM due to known under-counting in leaf count data.  Under-counting can be incorporated into the Bayesian model construction via inclusion of ``data uncertainty'' nodes or priors on node counts.  
We hypothesize that the true number of opioid overdose events in the province of British Columbia during the cohort period of 2015-2017 is significantly larger than the number of events captured by health administrative databases, as indicated by expert opinion regarding the possible magnitude of healthcare unattended overdoses.

\subsection{WMM Model Framework}
In the case that multiple subsets of the target population with known or estimated size are mutually exclusive, a tree can be constructed by defining the root to be the target population and the leaves to be sub-populations of the root.  Additional nodes along the root-to-leaf paths further describe nested subgroups, which may also be partially observed.  These structures have been used to described how subgroups are related to one another as well as the target population in a public health setting \cite{opioiddata}, and other applications in public health and epidemiology could similarly construct such ``data trees''.  The WMM exploits the tree structure to synthesize the partially observed evidence and generate an optimal estimate of target population given the constraints of the data.  Furthermore, since the traditional multiplier method is employed regularly in public health and epidemiology, the WMM may be more accessible than the application of fully Bayesian models via Markov chain Monte Carlo (MCMC) or Hamiltonian Markov chains (HMC). 

\begin{figure}
	\centering
	\includegraphics[width=.8\linewidth]{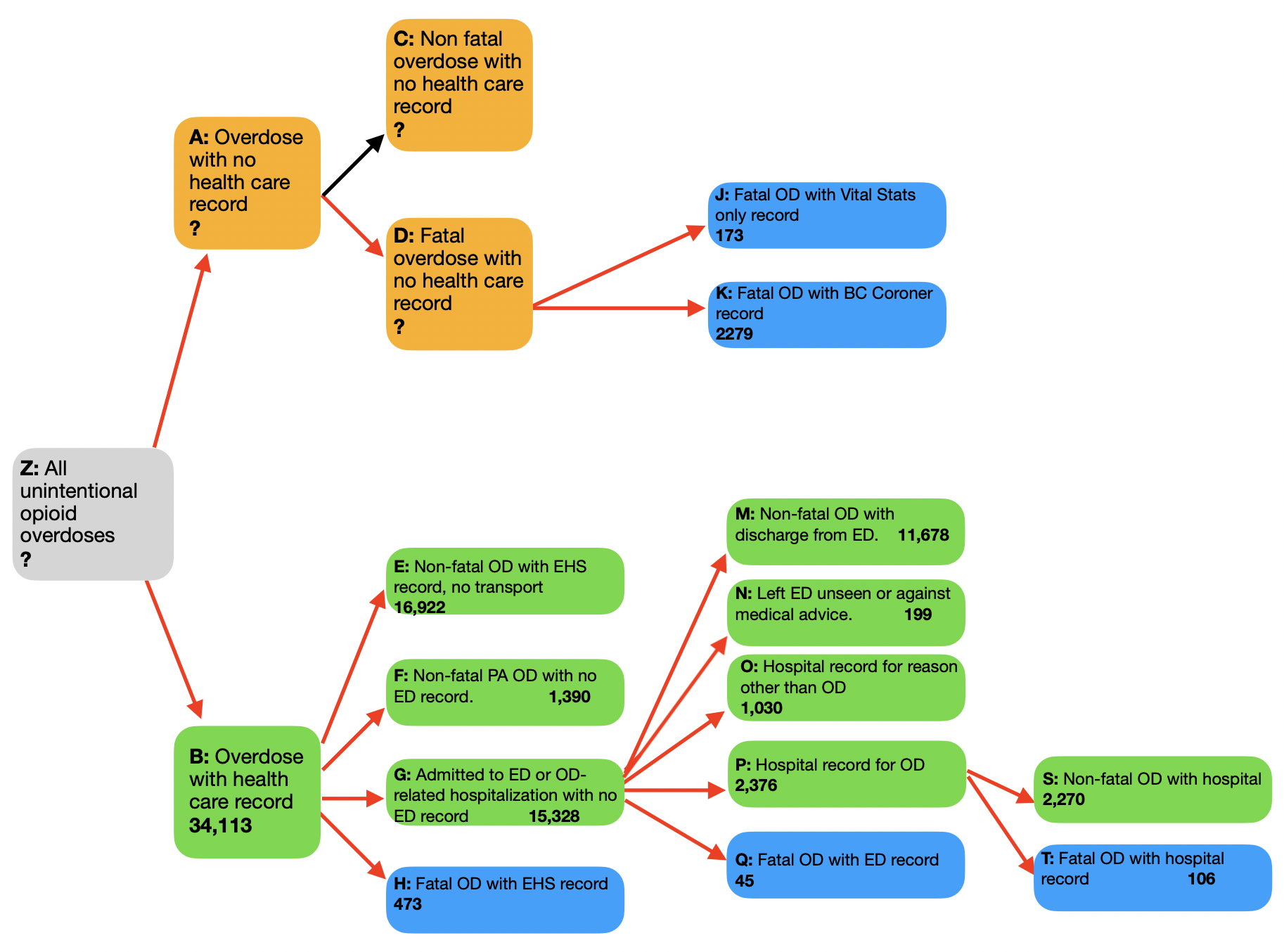}
	\caption{The full opioid pathways tree for WMM modeling;  ``full WMM modeling tree'', based on Figure \ref{fig:pathwaystree}.  Red paths indicate branching probabilities which must be informed by prior surveys or estimates to use the available leaf counts.}
	\label{fig:WMMtree}
\end{figure}

\begin{figure}
	\centering
	\includegraphics[width=.8\linewidth]{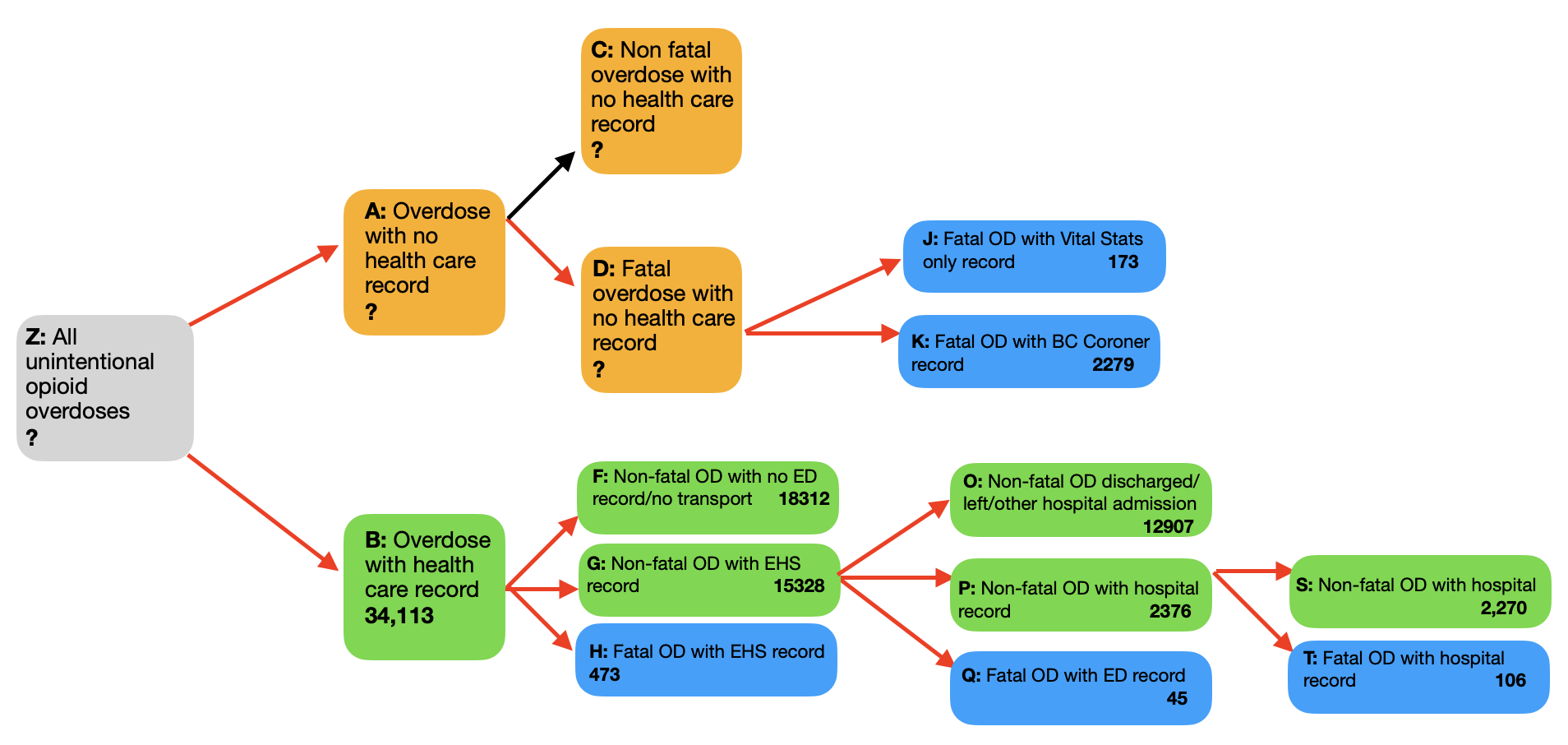}
	\caption{The ``simplified WMM modeling tree''.  Aggregates leaf nodes of Figure \ref{fig:WMMtree} to construct a simpler opioid pathways tree for WMM modeling, used for comparison to the full tree.  Red paths indicate branching probabilities which must be informed by prior surveys or estimates to use the available leaf counts.}
	\label{fig:WMMsimpletree}
\end{figure}

By referencing Figure \ref{fig:pathwaystree}, a modified pathways structure is constructed in Figure \ref{fig:WMMtree} which removes cyclic components by aggregating nodes and which further segregates healthcare-unattended deaths by stratifying this number by the source of data.
Let $\mathcal{T}$ denote the tree in Figure \ref{fig:WMMtree}.  Let $V^D(\mathcal{T})$ be the set of nodes of $\mathcal{T}$ for which we have data.  Furthermore, let $\mathcal{L}^*$ be the set of leaves, $L$, for which we require that both a marginal counts of $L$ is known and available branching estimates for all edges along the root-to-leaf path from $Z$ to $L$ are available:
\[
\mathcal{L}^* = \{E, F,H, J, K, M, N, O, Q, S, T\}.
\] 
Though multiple estimates exist for some branching, we use only one source of data to inform each branch.  Upper and mid-path node values are displayed in Figure \ref{fig:WMMtree}, however, these values not independently informed and represent only summations of lower leaves.  While these sums are not used to inform node counts, they are used to inform parameters of the $Beta$ and $Dirichlet$ branching distributions in the healthcare-attended arm of the tree.  In addition, some sibling leaf counts were generated by simply segregating data from the same source.  We also implement an alternative approach in which we opt for a simplified tree, as in Figure \ref{fig:WMMsimpletree}.
For this simplified tree, we have 
\[
\mathcal{L}^* = \{F,H,J,K,O,Q,S,T\}.
\]  
The healthcare unattended arm will use survey estimates or expert knowledge to inform the branching distributions and generate samples $p_{ZA}, p_{AD}, p_{DJ}, p_{DK}$.  The WMM procedure reflects data uncertainty only in the sampling of branching probabilities, and marginal counts are assumed exact in this application.

For the purpose of this analysis, back-calculations are made using branch distributions from a case-based sampling scheme which is decided by how many branches of a sibling group are informed, and whether or not they are informed by the same source of data; further details regarding this implementation can be found in Flynn and Gustafson \cite{flynncomp}.  
In short, $Beta(x+1, n-x+1)$ branch distributions are used where data sources differ among members of sibling branch group.  
For any branch between nodes $U$, $V$, with probability distribution $p_{UV} \sim Beta(x+1, n-x+1)$, the value $x$ represents the number of individuals in the informing sample survey who were counted at $V$, while the values $n$ are given by the total number of  individuals in the survey \cite{flynnmethods}.  Importance sampling ensures that independently chosen samples satisfy the constraint that the set of all sibling branch probabilities sums to 1.
Where one survey informs all members of the sibling group, or of a subset of a sibling group, $Dirichlet$ distributions are used.  Between these edge cases, a mixed approach using both $Dirichlet$ and $Beta$ distributions is used, in conjunction with importance sampling and rejection schemes \cite{flynncomp}.  This results in the distributions and parameter choices in Table \ref{table:WMMsimple} for Figure \ref{fig:WMMsimpletree}.  For the full tree, we have $p_{ZB}$, $p_{ZA}$, $p_{AC}$, $p_{AD}$, $p_{DJ}$, $p_{DK}$ as in Table \ref{table:WMMsimple}, as well as the additional branching specifications in Table \ref{table:WMMfull}. Notationally, $p_{xy}$ is used to denote the distribution of the random variable representing the branching probability along the directed path between nodes $x$ and $y$.  

\begin{table}[ht]
	\centering
	\begin{tabular}{|c|c|c|}
		\hline
		\rule[-1ex]{0pt}{2.5ex} Distribution & $x$ & $n$ \\
		\hline
		\hline
		\rule[-1ex]{0pt}{2.5ex} $p_{ZB}$ & 3 & 5 \\
		\rule[-1ex]{0pt}{2.5ex} $p_{ZA}$ & 2 & 5 \\
		\hline \hline
		\rule[-1ex]{0pt}{2.5ex} $p_{AD}$ & 1 & 10 \\
		\hline \hline
		\rule[-1ex]{0pt}{2.5ex} $p_{DJ}$ & 173 & 2452 \\
		\rule[-1ex]{0pt}{2.5ex} $p_{DK}$ & 2279 & 2452 \\
		\hline \hline
		\rule[-1ex]{0pt}{2.5ex} $p_{BF}$ & 18312 & 34113 \\
		\rule[-1ex]{0pt}{2.5ex} $p_{BG}$ & 15328 & 34113 \\
		\rule[-1ex]{0pt}{2.5ex} $p_{BH}$ & 473 & 34113 \\
		\hline \hline
		\rule[-1ex]{0pt}{2.5ex} $p_{GO}$ & 12907 & 15328 \\
		\rule[-1ex]{0pt}{2.5ex} $p_{GP}$ & 2376 & 15328 \\
		\rule[-1ex]{0pt}{2.5ex} $p_{GQ}$ & 45 & 15328 \\
		\hline \hline
		\rule[-1ex]{0pt}{2.5ex} $p_{PS}$ & 2270 & 2376 \\
		\rule[-1ex]{0pt}{2.5ex} $p_{PT}$ & 106 & 2376 \\
		\hline
	\end{tabular}
	\caption[Survey values for $Beta$ and $Dirichlet$ branch distributions of simplified pathways diagram, Figure \ref{fig:WMMsimpletree}.]{Survey values for $Beta$ and $Dirichlet$ branch distributions of simplified pathways diagram, Figure \ref{fig:WMMsimpletree}.}
	\label{table:WMMsimple}
\end{table}

\begin{table}[ht]
	\centering
	\begin{tabular}{|c|c|c|}
		\hline
		\rule[-1ex]{0pt}{2.5ex} Distribution & $x$ & $n$ \\
		\hline
		\hline
		\rule[-1ex]{0pt}{2.5ex} $p_{BE}$ & 16922 & 34113 \\
		\rule[-1ex]{0pt}{2.5ex} $p_{BF}$ & 1390 & 34113 \\
		\rule[-1ex]{0pt}{2.5ex} $p_{BG}$ & 15328 & 34113 \\
		\rule[-1ex]{0pt}{2.5ex} $p_{BH}$ & 473 & 34113 \\
		\hline \hline
		\rule[-1ex]{0pt}{2.5ex} $p_{GM}$ & 11678 & 15328 \\
		\rule[-1ex]{0pt}{2.5ex} $p_{GN}$ & 199 & 15328 \\
		\rule[-1ex]{0pt}{2.5ex} $p_{GO}$ & 1030 & 15328 \\
		\rule[-1ex]{0pt}{2.5ex} $p_{GP}$ & 2376 & 15328 \\
		\rule[-1ex]{0pt}{2.5ex} $p_{GQ}$ & 45 & 15328 \\
		\hline \hline
		\rule[-1ex]{0pt}{2.5ex} $p_{PS}$ & 2270 & 2376 \\
		\rule[-1ex]{0pt}{2.5ex} $p_{PT}$ & 106 & 2376 \\
		\hline
	\end{tabular}
	\caption[Survey values for $Beta$ and $Dirichlet$ branch distributions of full pathways diagram, Figure \ref{fig:WMMtree}.]{Survey values for $Beta$ and $Dirichlet$ branch distributions of full pathways diagram, Figure \ref{fig:WMMtree}.  Note that Figure \ref{fig:WMMtree} also requires $p_{ZB}$, $p_{ZA}$, $p_{AC}$, $p_{AD}$, $p_{DJ}$, $p_{DK}$ as in Table \ref{table:WMMsimple}, but the additional branching probabilities defined above correspond to different branches than those in Table \ref{table:WMMsimple} due to relabelling.}
	\label{table:WMMfull}
\end{table}
For both the full tree and the simplified tree, 10000 sampling iterations within the WMM were used.  The full code for the implementation of the WMM method to each of Figures \ref{fig:WMMtree} and \ref{fig:WMMsimpletree} can be found in the supplementary materials.  We note that since the branching distributions in the healthcare-attended branch of the tree are informed by POC data, these population-level parameters are not independent from the marginal counts of nodes in this segment of the tree, and thus the data structure assumed by the methodology is an approximation in this application \cite{flynnmethods}.

\subsection{Bayesian Model Framework}
A full Bayesian hierarchical model with evidence synthesis of the data contrasts estimates provided by the WMM, and extends back-calculated estimates by modeling biases in the individual data sources due to non-random sampling and non-sampling errors.  Bayesian approaches are inherently well-suited to problems that require synthesizing information across several data sources, enabling better integration of the data with expert opinion, thus we expect this model to produce more precise estimation \cite{irvinevarbayes2019,prevostHCV2015,mcdonaldflu2014}.  
The sensitivity of the posterior to other plausible priors is tested in Section \ref{sec:priorsens}, and the value of various types of information is explored more in the analysis of Section \ref{sec:VOI}.   
Focus has also been given to developing an adaptable Bayesian model for this data structure which could apply to similarly structured data elsewhere, including extending the overdose estimation application to other jurisdictions in Canada and internationally, where such rich datasets are not typically available.  It is possible that posterior estimates from the analysis of BC data may provide informative priors to other jurisdictions. 

\begin{figure}
	\centering
	\includegraphics[width=.8\linewidth]{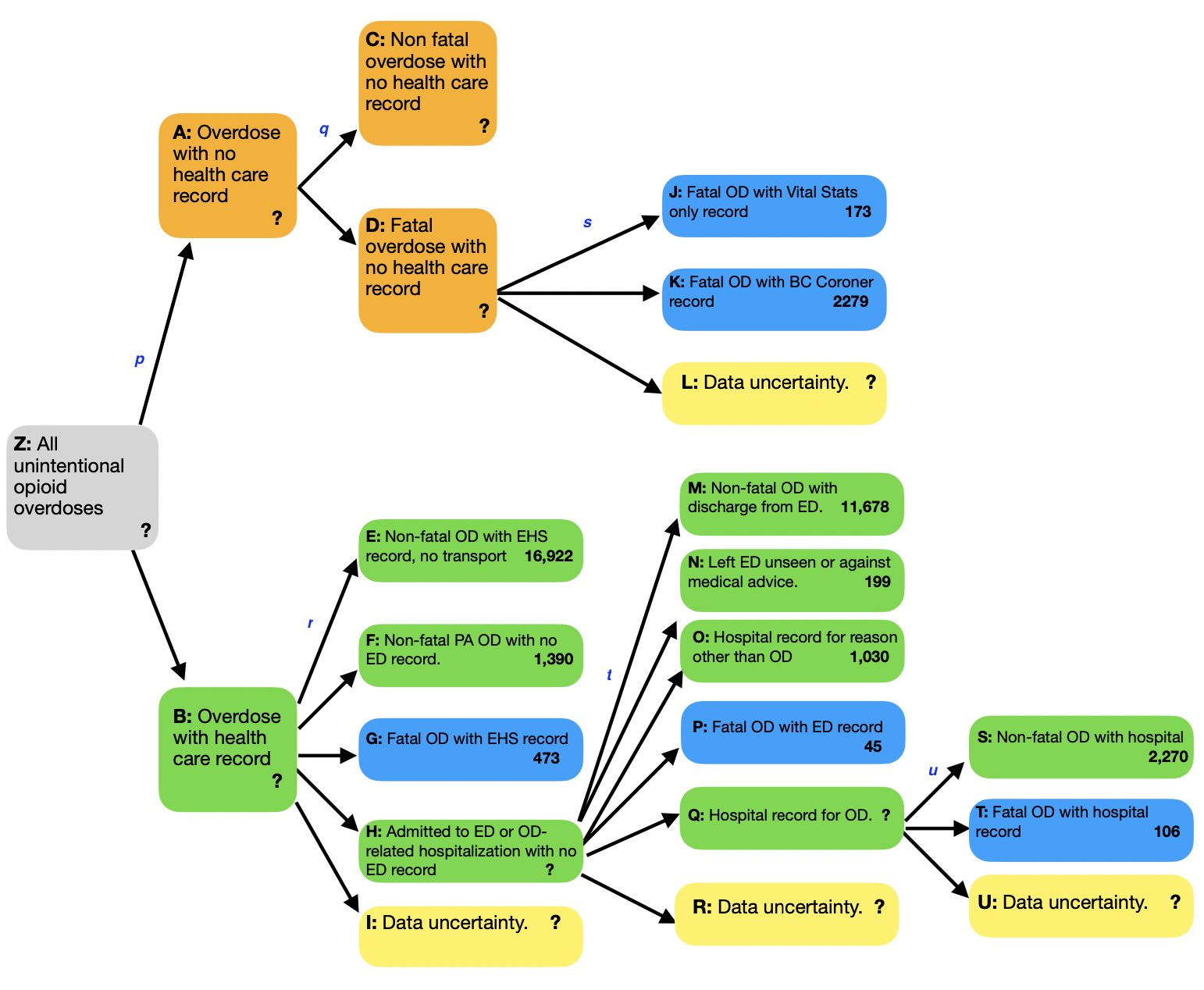}
	\caption{A modified version of Figure \ref{fig:pathwaystree}, used for the Bayesian model.}
	\label{fig:bayestree}
\end{figure}

\begin{figure}
	\centering
	\includegraphics[width=.8\linewidth]{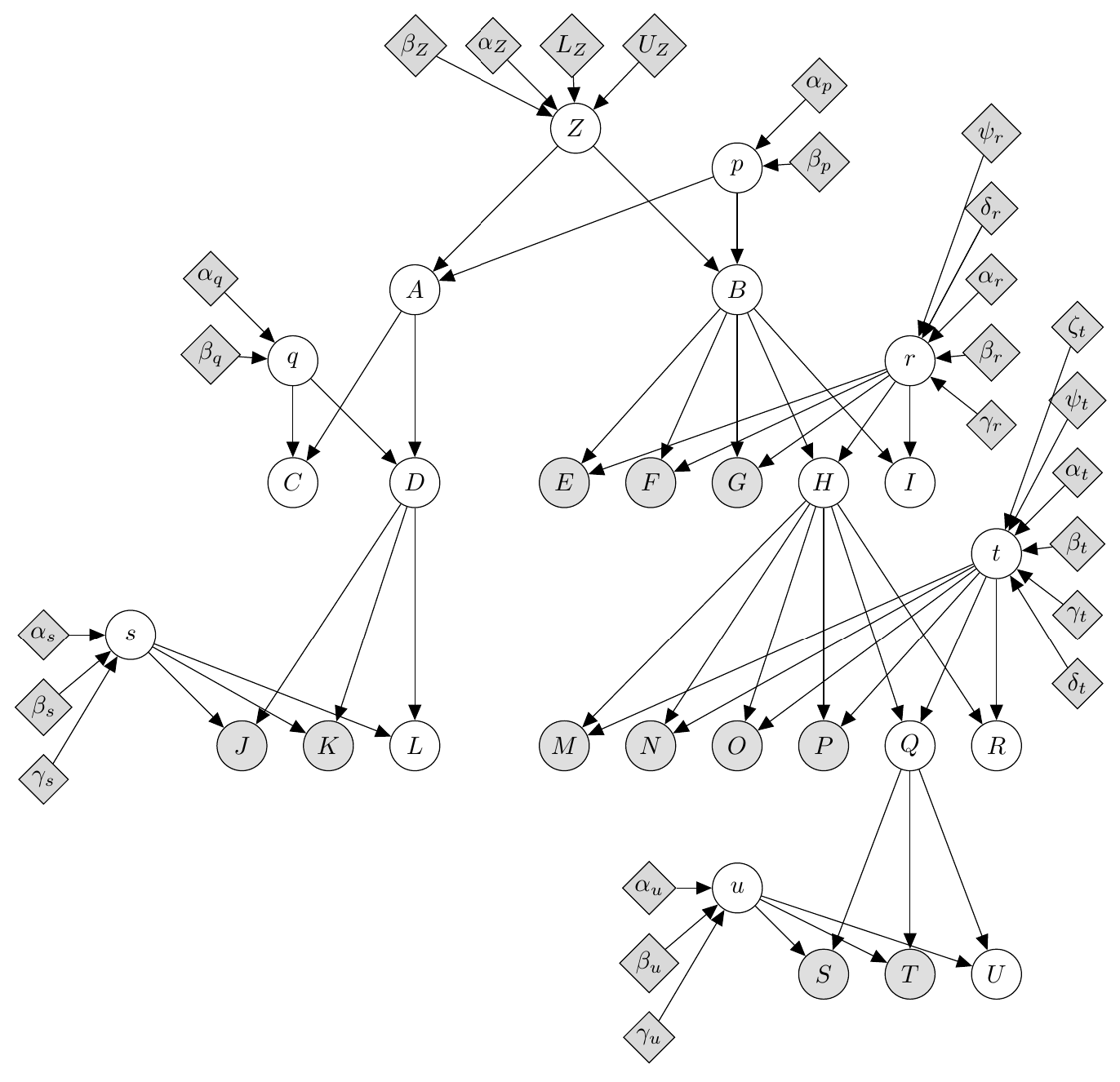}
	\caption{Graphical model for full tree Bayesian model, incorporating data errors and using latent parent nodes.  Shaded round nodes are observed, while unshaded round nodes are latent.  Diamond shaped shaded nodes are hyperparameters chosen using external prior knowledge such as literature estimates.}
	\label{fig:bayesDAG}
\end{figure}

The framework for a Bayesian model of the total number of opioid overdose events in BC from 2015-2017 is based on the tree of Figure \ref{fig:bayestree}.  This tree is akin to that used for the WMM in Figure \ref{fig:WMMtree}, but adds an additional node for data errors at all levels in which all leaves have marginal counts; these nodes incorporate the uncertainty in data collection and record-keeping. The model based on diagram Figure \ref{fig:bayestree} can be represented by the graphical model in Figure \ref{fig:bayesDAG}.  Parent or inner node values in Figures \ref{fig:pathwaystree} are ignored in the Bayesian model and \textit{not} treated as evidence, as these values are simply the upwards summation of values from observed leaves; including these observations would suggest data errors are equal to zero.  

Several priors must be chosen in this model.  We use Dirichlet distributions for all branching probabilities, which are only loosely informed by prior knowledge of field experts.  This is a fundamental difference between the sampling process of the Bayesian model and the WMM in this application, as the latter uses branch distributions curated from specific cohort data in many instances, and in addition, it samples some sibling branches using a mix of independent $Beta$ and $Dirichlet$ distributions with an importance sampling and rejection scheme.  Here, data uncertainties (nodes $L, I, R, U$) are informed by expert knowledge as a percentage of total possible events missed at each level.  This knowledge is translated to the prior by a choice of parameters in the corresponding Dirichlet distributions, which results in approximately that desired percentage of total individuals moving along the path to data uncertainty nodes.  The other parameters are set to be equal, so that uniform prior weight is assigned to all other branches at each level.  In particular, expert knowledge indicates that the populations at data uncertainty nodes $L,R,U$ is likely to be small (approximately 5-10\% of the parent population), with more certainty in this value in the healthcare-attended side of the tree.  Past data indicates that uncertainty at node $I$ may be closer to 15\% (the proportion of the population at node $B$ missed among the descendants), and there is comparatively more uncertainty in this estimate.  Thus we choose the following priors on branching probabilities, as defined by Figure \ref{fig:bayestree}:
\begin{itemize}
	\item $r \sim Dir(5, 5, 5, 5, 4)$
	\item $s \sim Dir(5, 5, 1)$
	\item $t \sim Dir(30, 30, 30, 30, 30, 12)$
	\item $u \sim Dir(30, 30, 5)$
\end{itemize}
As there are no data at any nodes in the branching at $p$ and $q$, we do not include latent data uncertainty nodes as children of parents $Z$ or $A$.  Prior knowledge obtained from administrative forms on bystander naloxone administration suggests node $A$ may represent approximately 40\% of the total population at $Z$ \cite{karamouzian2019}; furthermore, expert estimates indicate that 10\% of overdoses not attended by healthcare may result in fatalities (represented by node $D$).  These estimates direct our choice of parameters for branching $p,q$, and the following prior distributions are used:
\begin{itemize}
	\item $p \sim Dir(10,15)$
	\item $q \sim Dir(1,10)$
\end{itemize}

A $LogNormal$ prior is chosen for $Z$ and we incorporate postulated bounds on the range of the root node population size by choosing parameters of the distribution such that some percentage of the density is contained within these bounds.  These bounds are also expert specified with collaboration from PHAC, while considering a number of sources of data \cite{karamouzian2019,harmreductionsurvey,CCENDU,towardstheheart}.  A strict lower bound is given by the number of overdoses tracked within the BC cohort data, and an upper bound is chosen via an expert guided synthesis of the prior knowledge.
Thus the true value of $Z$ is estimated to be within $(34113,76621)$ with approximately 70\% certainty; practically, we aim to place roughly 70\% of the $LogNormal$ prior within these bounds by choosing a mean equal to a mid-range value and adjusting the standard deviation accordingly, resulting in the prior $Z \sim LogNormal(\log(51000), 0.38)$.  The MCMC modeling was performed using JAGS, with a JAGS model based on the graphical model of Figure \ref{fig:bayesDAG}.  The JAGS code was generated using the \textit{JAGStree} package in \texttt{R} \cite{r, flynncomp}.  After combining the data and the hyperparameters to generate the model input, the model is run using six chains, 10 million iterations, and 5 million iteration burn-in.

\section{Results}
\subsection{WMM Model Results}
\begin{figure}
	\centering
    \includegraphics[width=0.6\linewidth]{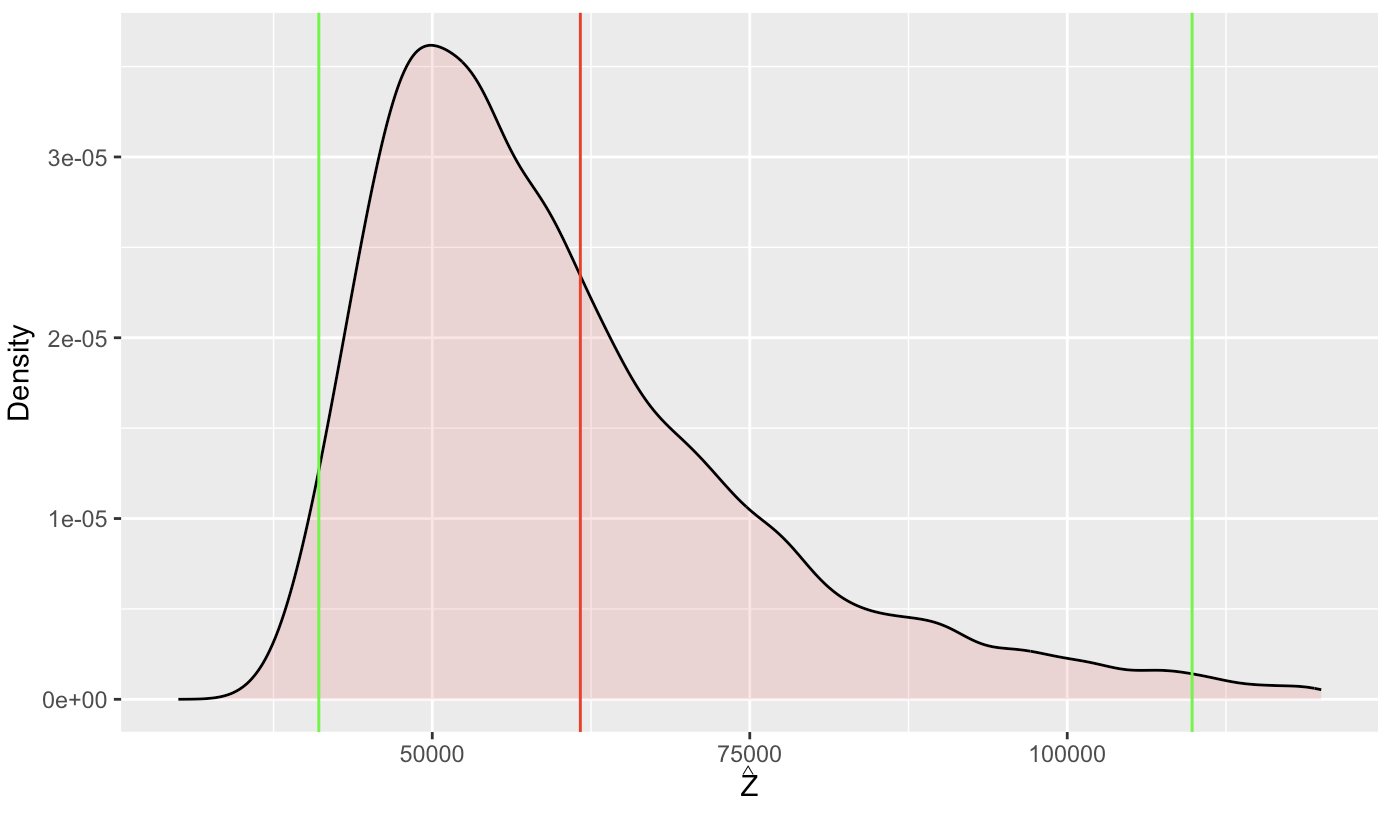}
	\caption{Distribution of the results of WMM applied to the full opioid tree, as in Figure \ref{fig:WMMtree}.  Vertical green lines represent the quantiles bounding the central 95\% of samples \cite{flynncomp}, and the red line is placed at the sample mean.}
	\label{fig:wmmopioiddistfull}
\end{figure}

\begin{figure}
	\centering
	\includegraphics[width=.6\linewidth]{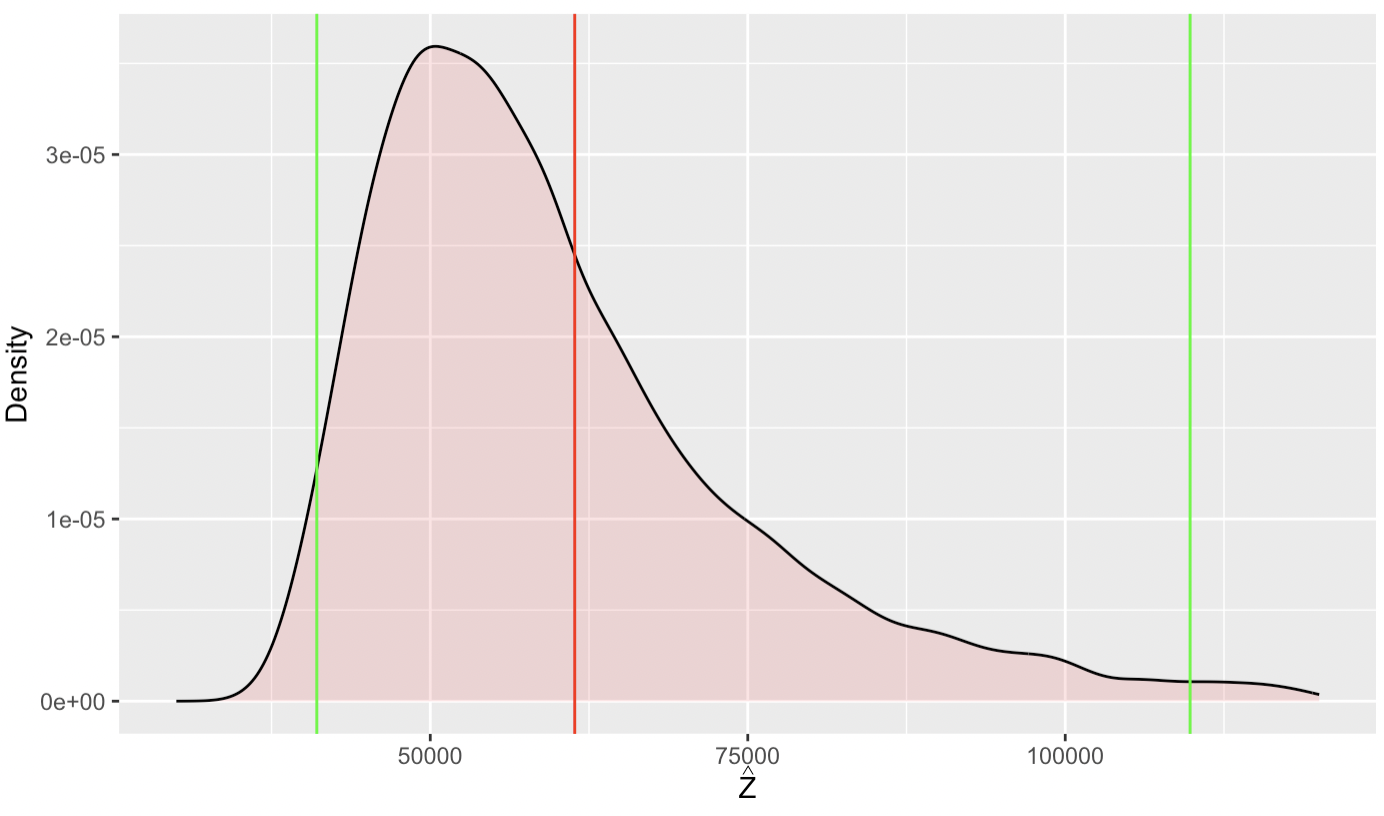}
	\caption{Distribution of the results of WMM applied to the simplified opioid tree, as in Figure \ref{fig:WMMsimpletree}.  Vertical green lines represent the quantiles bounding the inner of the 95\% of samples \cite{flynncomp} and the red line is placed at the sample mean.}
	\label{fig:wmmopioiddistsimple}
\end{figure}

The WMM was applied to both the full tree and the simplified tree (Figures \ref{fig:WMMtree} and \ref{fig:WMMsimpletree}, respectively).Two types of intervals were generated to quantify uncertainty: one which represents the central 95\% given by quantiles, and the 95\% confidence interval generated by the WMM function \cite{flynncomp}. The WMM applied to the full tree, as in Figure \ref{fig:WMMtree}, gives a mean estimate of $59445$ $(56815, 62196)$, and a median of $56845$ with central 95\% of the samples lying between $(41067,109830)$.  The distribution of results can be seen in Figure \ref{fig:wmmopioiddistfull}, and the weights the method has assigned to each root-to-leaf path can be found in Table \ref{table:opioidfullweights}.  For the simplified tree of Figure \ref{fig:WMMsimpletree}, we achieve similar results.  The estimated population size at node $Z$ is $59235$ $(56653, 61934)$, with a median of $56736$ and central 95\% of the samples between $(41146,110009)$.  The distribution of results is in Figure \ref{fig:wmmopioiddistsimple}, with the weights of each path in Table \ref{table:opioidsimpleweights}.  A slightly smaller confidence interval was observed when using the simplified tree with aggregate nodes.  As node aggregation results in larger $Dirichlet$ parameters and hence less variable distributions, smaller confidence intervals may be attributed to higher weighs assigned to aggregated paths.

\begin{table}
	\centering
	\begin{tabular}{|c|r|}
		\hline
		Path endpoint & Weight \\
		\hline \hline		
		J &  0.011\\
		K &  0.228\\
		E & 0.259\\
		F & 0.125\\
		M & -0.067\\
		N & 0.028\\
		O & 0.141\\
		S & 0.221\\
		T & 0.006\\
		Q & -0.012\\
		H & 0.060\\
		\hline 
	\end{tabular}
	\caption[Weights of each path of the full opioid tree (Figure \ref{fig:WMMtree}), as determined by the WMM. ]{Weights of each path of the full opioid tree (Figure \ref{fig:WMMtree}), as determined by the WMM.  The left column indicates the leaf which serves as the endpoint of a path from node $Z$, while the right column shows the weight assigned to this path by the model.}
	\label{table:opioidfullweights}
\end{table}

\begin{table}
	\centering
	\begin{tabular}{|c|r|}
		\hline
		Path endpoint & Weight \\
		\hline \hline		
		J &  0.021\\
		K &  0.210\\
		F & 0.484\\
		O & 0.056\\
		S & 0.175\\
		T & -0.025\\
		Q & -0.009\\
		H & 0.089\\
		\hline 
	\end{tabular}
	\caption[Weights of each root-to-leaf path of the simplified opioid tree (Figure \ref{fig:WMMsimpletree}), as determined by the WMM. ]{Weights of each path of the simplified opioid tree (Figure \ref{fig:WMMsimpletree}), as determined by the WMM.  The left column indicates the leaf which serves as the endpoint of a path from node $Z$, while the right column shows the weight assigned to this path by the model.}
	\label{table:opioidsimpleweights}
\end{table}

\subsection{Bayesian Modeling Results}
\begin{figure}
	\centering
	\includegraphics[width=.75\linewidth]{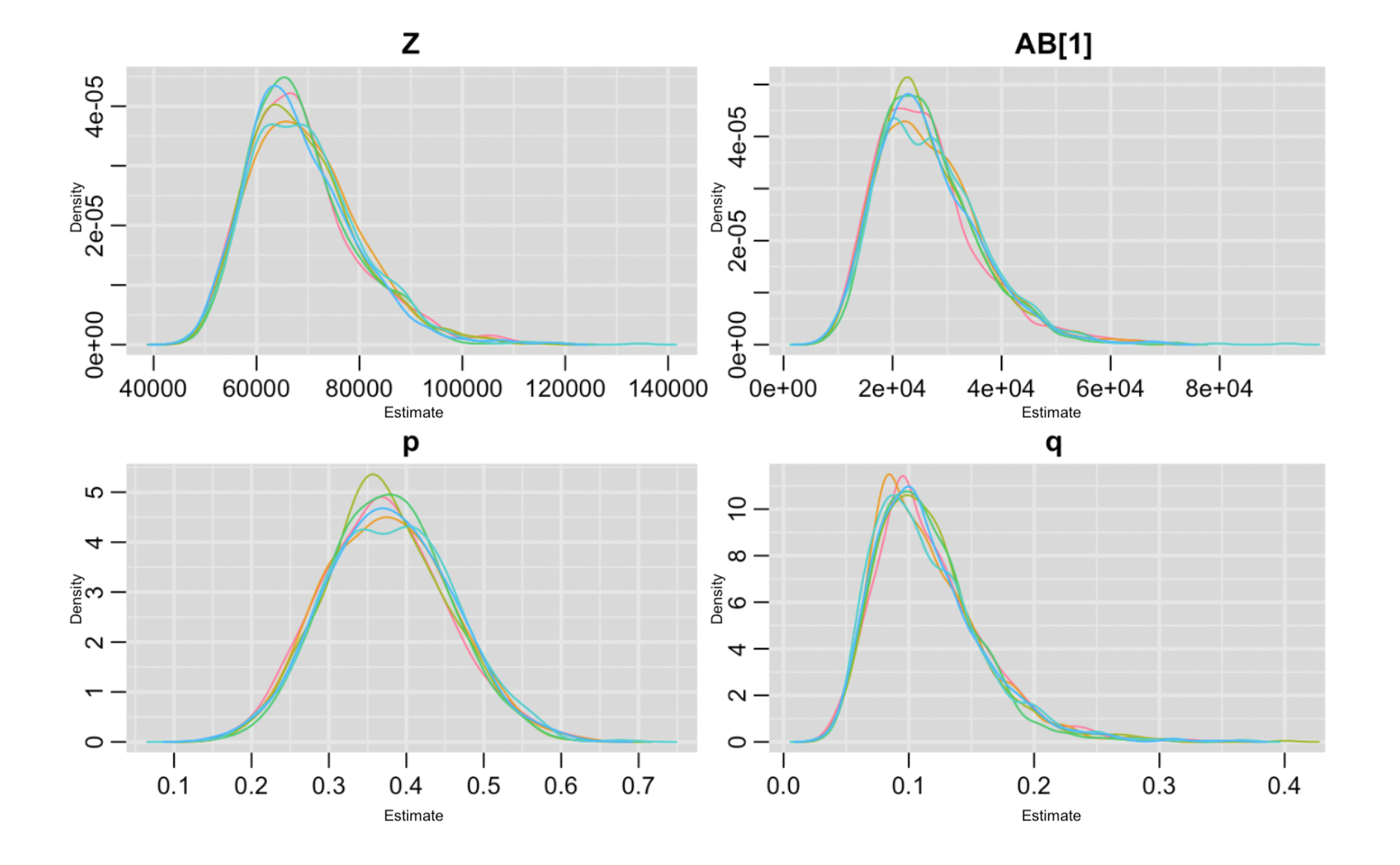}
	\caption[Posteriors of nodes $Z$, $A$, $p$, $q$ generated by the Bayesian model.] {Posterior distributions generated by the Bayesian model of nodes $Z$, the total number of opioid overdoses, $A$, the total number of healthcare-unattended overdoses, and probabilities $p$, the proportion of healthcare-unattended overdoses, and $q$, the proportion of overdose death in healthcare-unattended overdoses. Different coloured curves represent the six independent chains.}
	\label{fig:bayesdists}
\end{figure}

\begin{figure}
	\centering
	\begin{subfigure}
		\centering
		\includegraphics[width=.5\linewidth]{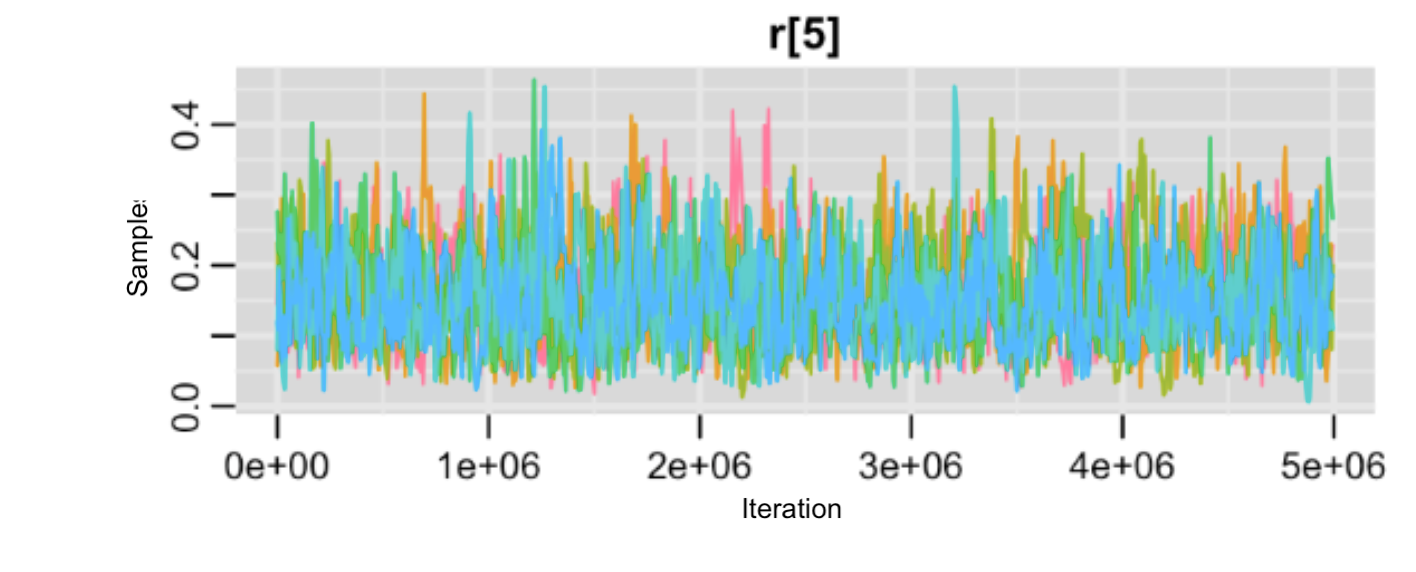}
	\end{subfigure}%
	\begin{subfigure}
		\centering
		\includegraphics[width=.5\linewidth]{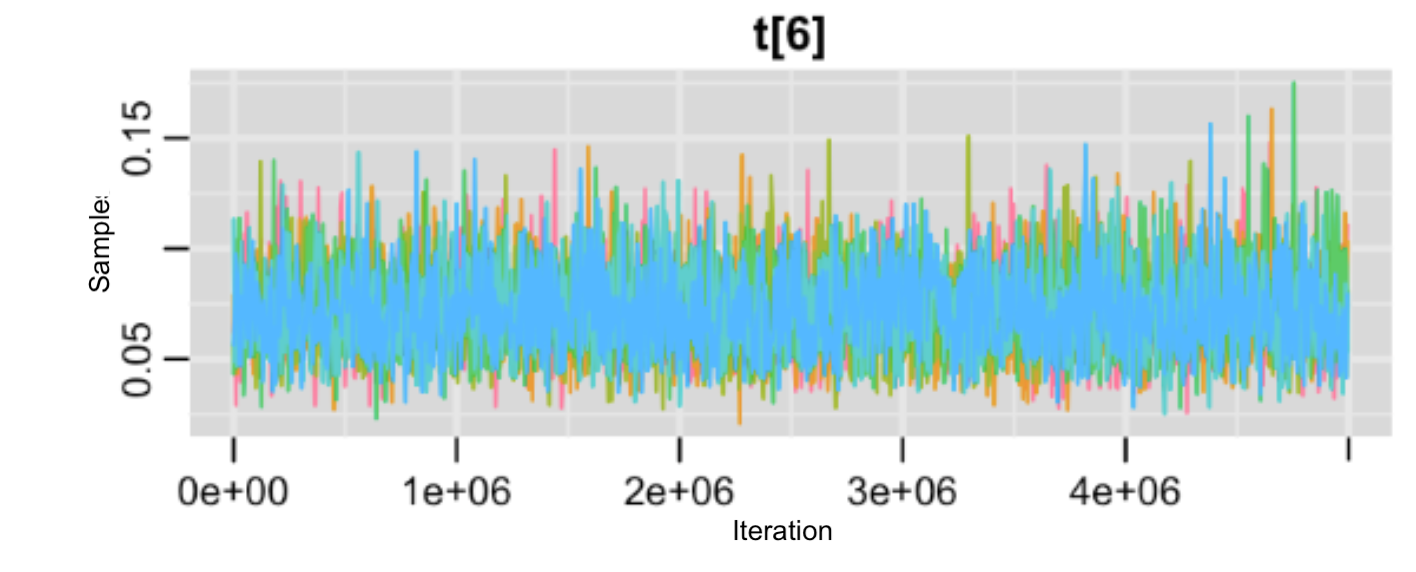}
	\end{subfigure}
	\caption{Trace plots from the Bayesian model on the full opioid tree of data uncertainty branching probabilities $r_I$, the proportion of uncounted overdoses among children of node $B$, and $t_R$, the proportion of uncounted overdoses among children of node $H$. Different colours represent the traces of the six independent chains.}
	\label{fig:bayestrace3}
\end{figure}

\begin{figure}
	\centering
	\begin{subfigure}
		\centering
    \includegraphics[width=.75\linewidth]{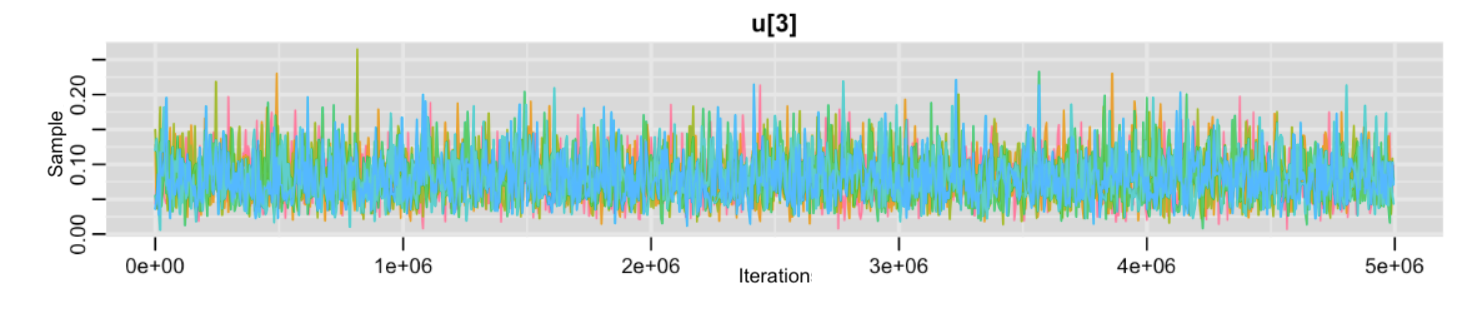}
	\end{subfigure}%
	\begin{subfigure}
		\centering
	\includegraphics[width=.75\linewidth]{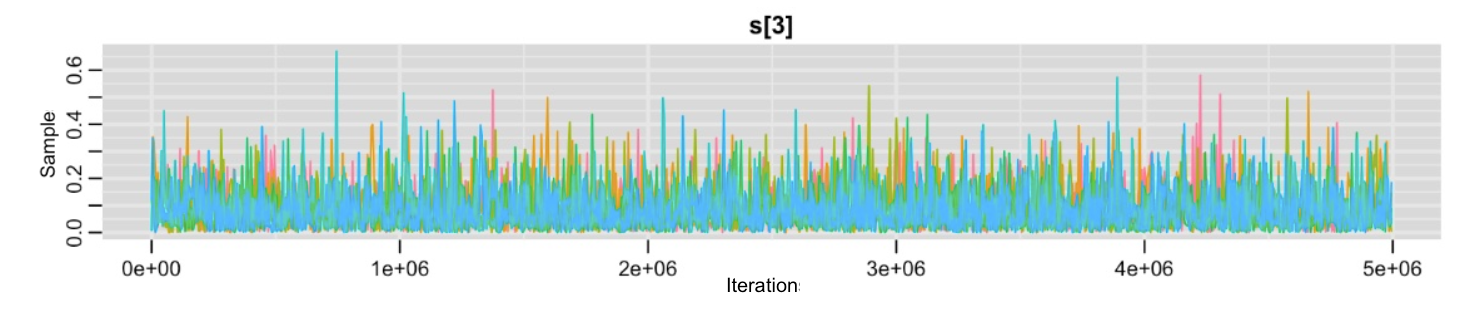}
	\end{subfigure}
	\caption[Trace plots of branches $u_U$ and $s_L$ from the Bayesian model.] {Trace plots from the Bayesian model on the full opioid tree of data uncertainty branching probabilities $u_U$, the proportion of uncounted overdoses among children of node $Q$, and $s_L$, the proportion of uncounted healthcare-unattended fatal overdoses which were not captured by Vital Statistics or BC Coroner data. Different colours represent the traces of the six independent chains.}
	\label{fig:bayestrace4}
\end{figure}

\begin{figure}
	\centering
	\includegraphics[width=.75\linewidth]{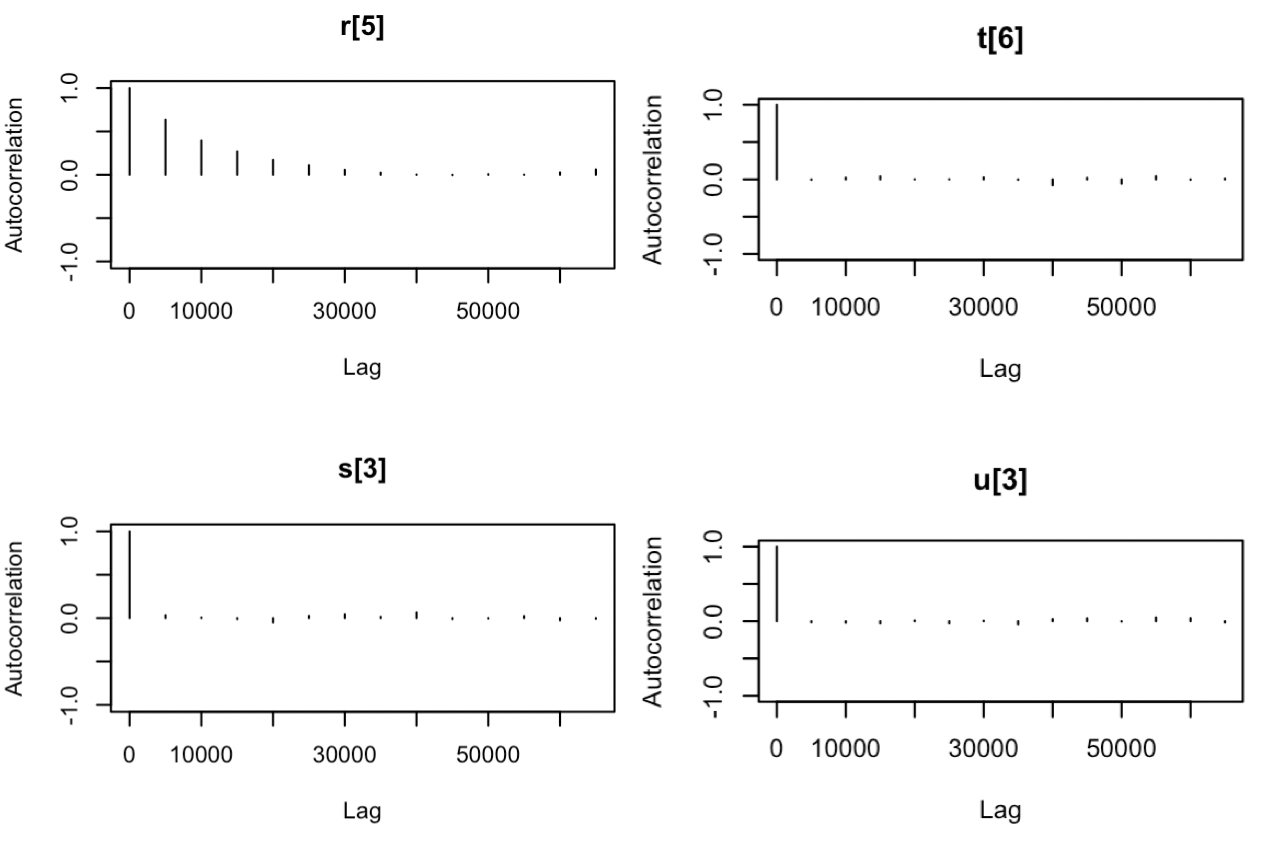}
	\caption[ACF plots of $r_I$, $t_R$, $u_U$, and $s_L$ from the Bayesian model.] {ACF plots of  branching probabilities $r_I$, $t_R$, $u_U$, and $s_L$ from the Bayesian model on the full opioid tree.}
	\label{fig:opioidACF2}
\end{figure}

The results of the Bayesian modeling can be found in Table \ref{table:bayesresults}, which summarizes point estimates, standard deviations, and 95\% credible intervals of parameters of interest.  The distributions of nodes $Z$ and $A$ can be found in Figure \ref{fig:bayesdists}, along with the posterior distributions of branching probabilities $p$, $q$. Trace plots of nodes $Z$ and $A$, branching probabilities $p$ and $q$, and the branching probabilities leading to data uncertainty nodes can be found in the supplementary (Figures \ref{fig:bayestrace1}, \ref{fig:bayestrace2}, \ref{fig:bayestrace3}, and \ref{fig:bayestrace4}). ACF plots of $Z$, $A$, $p$, and $q_D$ (the component of $q$ representing the proportion of events which move to node $D$) can be found in the supplementary, as Figure \ref{fig:opioidACF}, while ACF plots of $r_I$, $t_R$, $u_U$, and $s_L$ can be found in Figure \ref{fig:opioidACF2} in the same.  The effective sample sizes, $N_{eff}$ of the variables of interest can be found in Table \ref{table:opioidneff}.  These metrics suggest satisfactory mixing and convergence.

\begin{table}
	\centering
	\begin{tabular}{|c|c|c|c|c|}
		\hline
		Variable & Mean & SD & 2.5\% & 97.5\% \\
		\hline \hline		
		Z & 68978 & 10600 & 52634 & 93145\\
		A & 26585 & 9427 & 12465 & 48288\\
		B & 42394 & 3759 & 37245 & 51572\\
		C& 23865 & 9418 & 9868 & 45664\\
		I  & 6851 & 3732 & 1783 & 15915 \\
		L & 267 & 310 & 5 & 1100\\
		R & 1215 & 364 & 622 & 2016\\
		U & 214 & 101 & 66 & 456 \\
		\hline \hline
		$p$ & 0.376 & 0.079 & 0.229 & 0.537 \\
		$q_D$ & 0.115 & 0.043 & 0.054 &  0.220 \\
		$s_L$& 0.089 & 0.083 & 0.002 & 0.310\\
		$r_I$ & 0.156 & 0.069 & 0.047 & 0.310\\
		$t_R$ & 0.072 & 0.020 & 0.038 & 0.114\\
		$u_U$ & 0.081 & 0.034 & 0.027 & 0.160\\
		\hline
	\end{tabular}
	\caption[Results of interest in the full overdose pathways Bayesian model. ]{Posterior means, standard deviations, and credible intervals of the parameters of interest in the full overdose pathways Bayesian model.}
	\label{table:bayesresults}
\end{table}

\begin{table}
	\centering
	\begin{tabular}{|c|c|}
		\hline
		Variable & $N_{eff}$ \\
		\hline \hline		
		Z & 2300 \\
		A & 5800 \\
		B & 1400 \\
		C& 6000 \\
		I  & 2000 \\
		L & 6000 \\
		R & 6000 \\
		U & 3800 \\
		\hline \hline
		$p$ & 6000 \\
		$q_D$ & 6000 \\
		$s_L$& 4900 \\
		$r_I$ & 2200 \\
		$t_R$ & 6000 \\
		$u_U$ & 4200 \\
		\hline
	\end{tabular}
	\caption[Effective sample sizes for the full overdose pathways Bayesian model. ]{Effective sample sizes for the parameters of interest in the full overdose pathways Bayesian model.}
	\label{table:opioidneff}
\end{table}

The model estimates the total number of overdoses, including both healthcare-attended or -unattended, to be approximately 68978 (52634, 93145).  The majority of the uncertainty in this estimate results from the estimate at node $A$, the total number of healthcare-unattended overdoses, and more specifically, node $C$, the number of non-fatal overdoses in this pathway.  The model estimates that the total number of unattended overdoses at node $A$ to be approximately 26585 (12465, 48288).  This includes 23865 (9868, 45664) survived overdoses at node $C$, or 89.8\% of all unattended events.  A further 267 (5, 1100) uncounted deaths were estimated at node $L$, or 9.8\% of all unattended deaths.

The total number of healthcare-attended overdoses at node $B$ is estimated at 42394 (37245, 51572), compared to the 34113 confirmed overdoses within the POC dataset.  This discrepancy results from the data uncertainty nodes introduced in the healthcare arm of the pathways tree.  The model estimates 16.2\% of the overdoses at node $B$ were uncounted at node $I$, or 6851 (1783, 15915).  At node $H$, a further 7.3\% of events were estimated uncounted (node $R$), or 1215 (622, 2016).  An additional 8.3\% of overdose events from node $Q$ were estimated uncounted and captured by node $U$, or 214 (66, 456).

\subsection{Validation and Model Fit}
\subsubsection{Prior Sensitivity}\label{sec:priorsens}
Other reasonable priors informed by past literature and surveys were used to adjust priors on $Z$, $p$, or $q$, and determine model sensitivity.  For the prior on $Z$, we adjust the soft bounds informing the parameters of the $LogNormal$ distribution, the certainty in these bounds, and also try a uniform prior on $Z$. For $p$, the probability that an overdose is unattended by healthcare, we explore the model's sensitivity to this prior by adjusting the $Beta$ parameters, as information around this branching is not derived from health administrative data and is highly uncertain relative to the healthcare-attended arm of the tree.  Furthermore, being a branch that emanates from the root node, it likely an important modeling component as it affects all paths below it.
Lastly, we adjust the prior distribution $f(q)$, where $q$ is the probability that an unattended overdose results in death. 

When confidence in the soft bounds of $Z$ remains fixed but tighter bounds are chosen, model estimates of $Z$ and $A$ decrease.  The change in $A$ is primarily due to a decrease in the estimate of $C$, the number of healthcare-unattended survived overdoses.  The number of healthcare-attended overdose events estimated at node $B$ is also slightly lower, but the majority of the model adjustment being reflected in the healthcare-unattended arm of the tree (where less data is available), and in particular, there is also a significant decrease in the mean estimate of $p$.  In general, we do not expect the population at node $B$ to vary drastically between any of the scenarios explored during sensitivity analysis, as it is informed by many sources of data being used together to influence its estimated value.

When the $Z$ prior is changed to reflect an increase in the confidence in soft bounds to 95\%, model estimates of $Z$ and $A$ again decrease.  This is unsurprising, as decreased prior variability will result in fewer extreme samples.  The change is $A$ is again primarily explained by a decrease in the estimate of $C$, though the estimates of branching probabilities $p$ and $q$ are remain relatively stable.
By contrast, when the confidence in soft bounds is decreased, the estimates of $Z$ and $A$ increase, driven primarily by an increase in the events at node $C$, the number of non-fatal unattended overdoses.  The proportion of fatal unattended fatal incidents, $q$, is estimated to be only slightly lower, while the proportion of all unattended overdoses, $p$, is estimated to be only slightly higher than the original model, though with wider credible intervals.

For a model with a uniform prior on $Z$, increasing iterations was necessary for convergence, though the estimates of nodes $Z$ and $A$ were only marginally smaller, while $p$ and $q$ estimates were stable.

When the prior on $p$ is adjusted, large deviations from the original model estimates of $Z$, $A$, and $C$ were observed.  When the prior on $p$ is adjusted to reflect the upper bound of prior knowledge of this parameter, $Z$ is estimated to be approximately 70\% larger than in the original model, with the difference again primarily attributed to an increase at node $A$, and a corresponding increase in the estimated probability of unattended overdose, $p$. The proportion of non-fatal overdoses is also estimated to be smaller; this is unsurprising, as the number of survived overdoses is unknown.
Analogous opposing changes are observed when the prior on $p$ is adjusted to reflect the lower bound of prior knowledge on this parameter.  
When a uniform prior on $p$ is used instead, posterior estimates match the original model very closely.

When a uniform prior is used on $q$, the proportion of unattended fatal incidents, the posterior of $q$ converges and the mean estimate is slightly higher than in the original model, however, there is not large difference in the other parameters of interest.

A summary of posterior estimates under each of the adjusted priors can be found in the supplementary.

\subsection{Value of Information}\label{sec:VOI}
Significant effort on the behalves of the BCCDC and PHAC was put into the creation of the linked POC dataset.  Other jurisdictions may not have these resources available, thus determining robustness of models with respect to missing data sources or simplified data structure is of interest for wider applicability.  
Deleting data enables us to examine what we can expect from the model when certain nodes are unknown.  This may be especially interesting on the healthcare-unattended side of the tree, where it is likely that information is scant or lacking for most jurisdictions.  We expect the accuracy of estimation to be heavily dependent on which data are lacking, the position in the tree, and the relative proportion of knowledge that the deleted data represents to the local structure.  In addition to data deletion, it is of interest to determine if such a segregated tree structure is required for accurate estimation, since aggregate data may be more readily available and carry less privacy concerns than segregated data.  In the Bayesian modeling, past results indicate that aggregate nodes should generate similar estimates \cite{flynnmethods}.  Furthermore, for the POC data, the segregated and aggregated trees produce similar results with the WMM, suggesting that placing significant effort into the stratification of data may not be worthwhile when employing the WMM for root node population estimation.  Here, we now perform the same experiment with the Bayesian model, aggregating the data of all leaf nodes of the same level into one terminal node at that level, and observing the effect on parameter estimates.

We remove the data and re-run the model for nodes J (fatal healthcare-unattended overdoses recorded only in Vital Statistics), node K (fatal healthcare-unattended overdoses with BC Coroners record), node S (non-fatal healthcare-attended overdose with hospital record; one of two lowest-level nodes), both S and T (fatal healthcare-attended overdose with hospital record; both lowest-level nodes), both J and K (which represent all count data in the healthcare-unattended segment of the tree), and all fatality nodes (blue nodes of Figure \ref{fig:bayestree}; complete deletion of healthcare-unattended nodes and no fatality knowledge).  

The most significant changes to posterior estimates occur locally; nodes or branches at shorter distance from those with deleted data are affected by the data omission, with some transfer upstream to estimates of $Z$, $A$, and $B$.  The MCMC are took more iterations to converge, which is to be expected.  The observed changes in posterior estimate were generally small, with the exception of the case when all fatality node data is omitted.  The total count at fatality nodes represents only a small proportion of all observed overdose events, as most data is obtained from the healthcare-attended setting where the vast majority of overdoses are survived.  However, the deletion of these data introduces an unknown node at multiple levels of the tree within both major branches (healthcare-attended and -unattended).  The branching priors place uniform weight on all sibling nodes that are not \textit{data uncertainty} nodes, so in the case that all fatality data is removed from the model, no evidence is being used to indicate that the values at these nodes may by lesser than their siblings.  Without this knowledge, the model grossly overestimates the values at these nodes, increasing the overall estimate of $B$ and thus $Z$.  By contrast, the WMM simply omits paths without counts at the leaves; thus the deletion of count data did not significantly affect WMM estimates.

\subsection{WMM Model Sensitivity Analysis}\label{sec:wmmsens}
While the WMM does not make use of prior distributions, a sensitivity analysis can similarly be conducted on the branching estimates of the WMM model.  In particular, an assumption has been made that $p \approx 40\%$ in order to inform the choice of distribution parameters in the Bayesian model, and similarly for the distribution parameters of $p_{ZA}$ and $p_{AD}$ in the WMM model.  
Similar results from a sensitivity analysis on branching distributions is observed with the WMM model as for the Bayesian model, though the WMM shows greater sensitivity to alternate priors for $p_{ZA}$; estimates of the root node population are more drastically affected by adjustments in branching distribution parameters.  This may be partially attributed to a greater synthesis of available evidence in the Bayesian model, which ultimately converges on a posterior branching distribution which also considers the likelihood of the evidence in the tree structure.

We also adjust the distribution on $p_{AD}$, the proportion of unattended overdose fatalities.  Shifting the mean of the density of this parameter to higher values causes an overall decrease in the estimate of node $Z$, which is as expected, and was also observed in the Bayesian model.  However, we again observe a more drastic change in the estimate of node $Z$ when adjusting this distribution, demonstrating the greater sensitivity of the WMM model as compared to the Bayesian model.  A summary of estimates under each adjusted distribution can be found in the supplementary material.

\section{Discussion}
The Bayesian model estimated the population at node $Z$ to be 68978 $(52634, 93145)$, while the WMM estimated this same value to be 59445 $(56815,62196)$.  The value estimated by the Bayesian model fell outside the confidence interval produced by the WMM; the WMM also produced a much tighter confidence interval than the credible interval of the Bayesian model.  There are several likely factors attributed to these differences.  The construction of $Beta$ parameters in this application uses cohort (population level) values, which resulted in large parameters and correspondingly narrow distributions.  Since the majority of the weight in the WMM was placed on healthcare-attended paths, we expect relatively low variability while using the standard confidence interval of the WMM, making the interval unreliable in this context \cite{flynncomp}.  The central 95\% interval using quantiles may be more reliable for this setting.  In addition, the discrepancy between the estimates generated by the two methods can explained by the estimated number of missed events - the Bayesian model estimates that more than 8000 overdose events are missed within the healthcare-attended branch of the tree, approximately equal to the difference between the two model estimates.  This indicates that the treatment of node counts as fixed values in the WMM may be a limiting construction when node counts are expected to be biased, and that a extension of the WMM methodology which includes distributions on node counts could significantly improve estimate accuracy in some circumstances.  
Since errors in node counts are known to exist in this data, all sources of uncertainty are not accounted for by the WMM, so a model which assigns distributions to these nodes would be beneficial.  Alternatively, researchers may choose to augment data by increasing total leaf node counts by an estimated number of missed events.  In any case, the lack of accounting for node count uncertainty suggests that confidence interval produced by the WMM model will be artificially narrow in this setting.  

The sensitivity analysis of the Bayesian model indicated that the inference was significantly affected by the choice of priors on $Z$ and $p$.  
In general, where a new prior on $Z$ resulted in lower posterior estimates of $Z$, this could be explained primarily through a higher posterior mean in the proportion of healthcare-unattended fatalities, $q$.  Similarly, when estimates of $Z$ were higher, the posterior mean of $p$ remained relatively stable, but the proportion of fatal/non-fatal healthcare-unattended overdoses reflected the majority of the change, ultimately seen in an increase in the population at node $C$, non-fatal healthcare-unattended events. Posterior estimates were more sensitive to changing branching priors nearer to the root of the tree.

The WMM model analysis also revealed sensitivity to the branching distributions, again, especially to those branches closest to the root.  This is unsurprising, as branching closer to the root affects a larger number of back-calculated estimates.  Thus it is important that prior knowledge used to inform branching in this model is carefully chosen, especially for those values closer to the root.  This method is at an inherent disadvantage to Bayesian modeling with respect to branching distributions; though conflicting information among sibling branches is mitigated by the rejection scheme, branching distributions are not updated based on the evidence (as a prior is) and are thus limited to the original inputs.   

In evaluating the value of information, the Bayesian model was robust in most scenarios of missing data.  The largest differences in parameter estimates were observed locally, while upstream parameter estimates were relatively stable.  An exception occurred when all fatality data were omitted and uniform branching priors were still used.  This demonstrates the sensitivity of the Bayesian model to branching distributions in the absence of count data; uninformative priors may not be appropriate in the absence of node data along a given path. 

Segregation of nodes does not greatly affect the root node estimates in either model.  In the Bayesian model with aggregated nodes, the mean estimate of $Z$ is 68934 $(52983,93379)$ as compared to the estimate of 68978 $(52634, 93145)$ with the fully segregated model.  Similarly, the full WMM model produces a $Z$ estimate of 59445 $(56815, 62196)$ as compared to the simplified tree with aggregated nodes, which generates a mean estimate of $Z$ at 59235 $(56653, 61934)$.  These results suggest that in some cases, there may not be a worthwhile improvement from further refining data.  In addition, if resources are limited and segregated data is unavailable, researchers should not be discouraged when using either method to generate root node population estimates if sufficient aggregate data is available.  While both the WMM and the Bayesian models showed robustness to aggregated node counts, we reiterate that model construction and segregation/aggregation should be carefully considered.  For example, deleting data at nodes $J$ and $K$ in the Bayesian model showed that while the model was robust to estimating an aggregate total for these nodes, there was insufficient local knowledge to appropriately estimate each node individually.  Furthermore, robustness of model estimates of $Z$ were not symmetric to sibling data deletion.  These results suggest that a general recommendation as to which nodes should be segregated or could be aggregated is not suitable, and instead depends largely on the specific use case and missing data.

BC's POC dataset is a linked dataset, rich in information.  Future work with this application could include stratification by age group, health authority, or sex (gender is not an available feature of the data), which could easily be examined using either model \cite{springergender2012}.  It would also be of interest to extend this analysis to other jurisdictions in Canada, where posterior distributions of $Z$ and $p$ could be used as priors to improve analysis if other provinces or regions have comparable conditions to those in BC during the 2015-17 time period.  
Additionally, while BC has been dealing with rising overdose rates since 2012, data from the COVID-19 pandemic period suggests this longer-standing epidemic rapidly worsened in the interplay between the two public health crises \cite{overdoses2022}.  Take-home naloxone kit intervention analysis has previously been conducted in BC \cite{irvinethnkits2018, irvinevarbayes2019}; however, evaluating the potential effectiveness of interventions such as this is crucial to establish an appropriate and effective response.

\section{Conclusion}
A WMM model and a hierarchical Bayesian model were constructed and implemented on the POC data in British Columbia, for the time period of 2015-2017.  The results of both models suggest a large number of overdose events are unaccounted for in the sources of data commonly used to inform the total number of overdoses, and there may be over 70\% more events occurring than the number suggested by aggregating raw healthcare administrative data alone.  It is important that health institutions and government bodies consider both healthcare-attended and -unattended events by using estimation techniques to uncover the true size of this hidden population, to allocate sufficient resources and plan more suitable interventions or services. 

The analyses on the POC data support the use of the WMM method as an alternative to a hierarchical Bayesian model, particularly in the absence of resources or expertise to implement a fully Bayesian approach and interpret convergence and output of such a model.  While the sensitivity of the WMM to the accuracy of branching estimates and node counts must be carefully considered, the WMM is easy to understand and interpret, and implementation of this method is simplified by using the \textit{AutoWMM} package in \texttt{R} \cite{flynncomp}.  

\subsection*{Declaration of Interests}
The authors have no conflicts of interest to declare.

\subsection*{Acknowledgments}
The authors would like to acknowledge the contributions of Margot Kuo, MPH, at the Public Health Agency of Canada, who provided critical expert prior knowledge of the existing literature, the data sources used for analysis, and the potential biases of these, from project conception to completion. 
We also acknowledge the contributions of Heather Orpana, PhD, and Eva Graham, PhD, also of the Public Health Agency of Canada, who provided their expert opinions and considerations from an epidemiological and public health perspective.  
This work was supported by NSERC Discovery Grant (RGPIN-2019-03957), an NSERC CGS-D, and a CIHR Doctoral Health System Impact Fellowship in partnership with the Public Health Agency of Canada (PHAC).

\clearpage
\section{Supplementary}

\begin{center}
	\includegraphics[width=.75\linewidth]{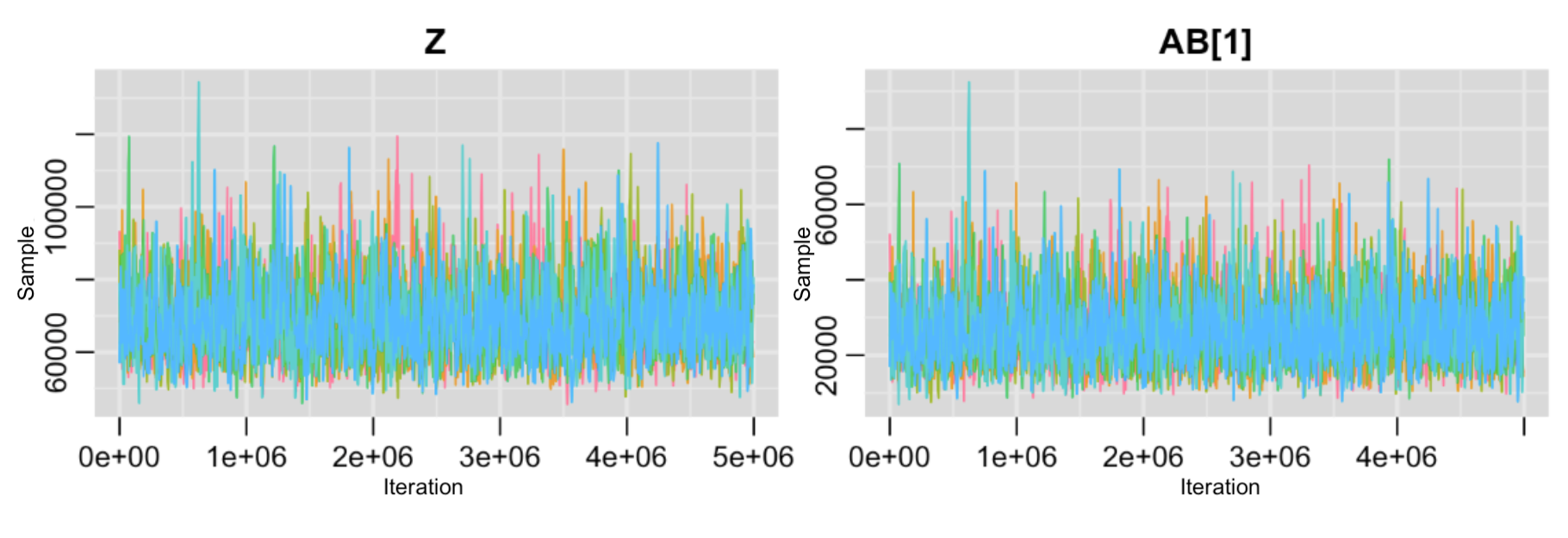}
	\captionof{figure}{Trace plots of nodes $Z$, the total number of opioid overdoses, and $A$, the number of healthcare-unattended overdoses, from the Bayesian model on the full opioid tree. Different colours represent the traces of the six independent chains.}
	\label{fig:bayestrace1}
\end{center}

\begin{center}
	\includegraphics[width=.75\linewidth]{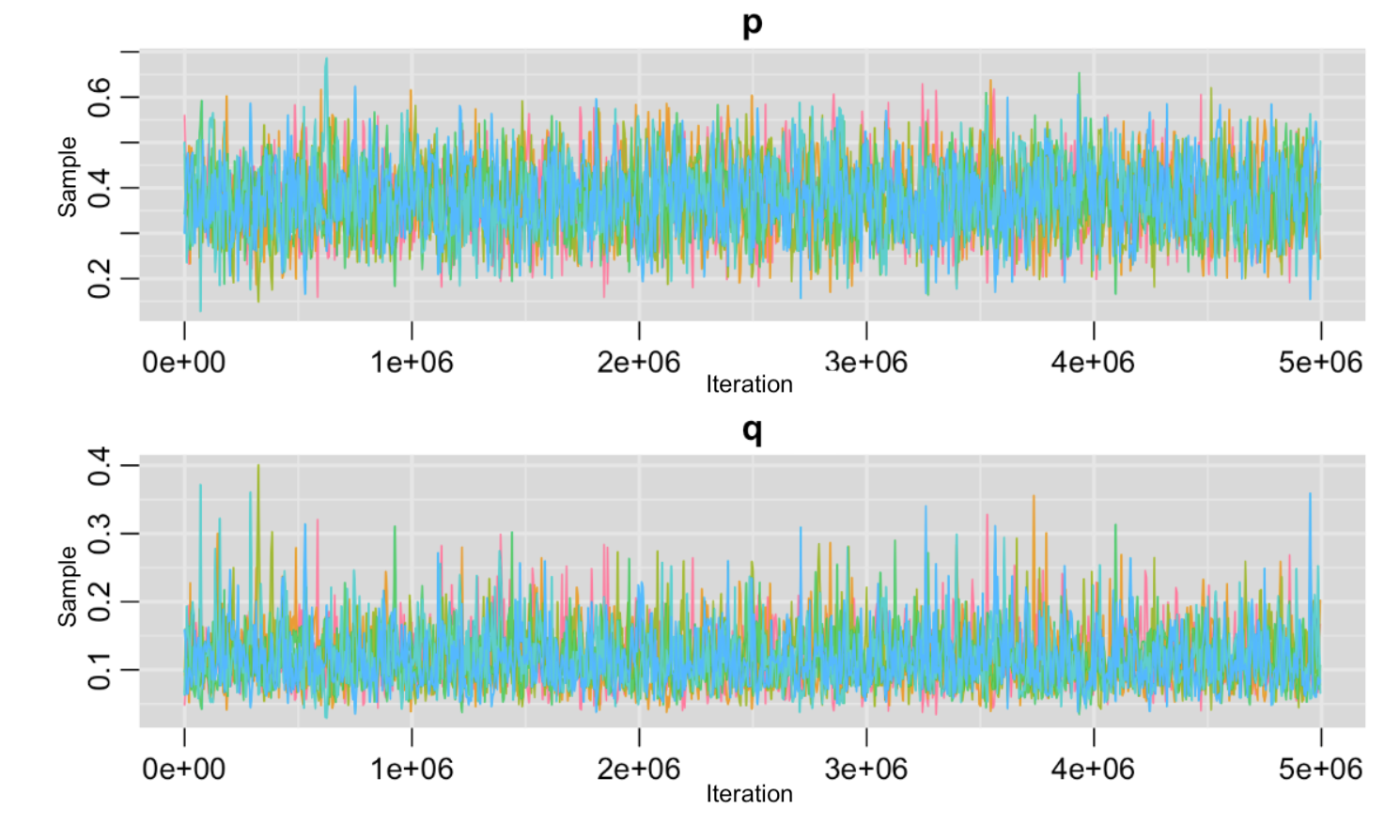}
	\captionof{figure}{Trace plots of branching probabilities $p$, the proportion of healthcare-unattended overdoses, and $q$, the proportion of healthcare-unattended overdose deaths from the Bayesian model on the full opioid tree.  Different colours represent the traces of the six independent chains.}
	\label{fig:bayestrace2}
\end{center}

\begin{figure}
	\centering
	\includegraphics[width=.75\linewidth]{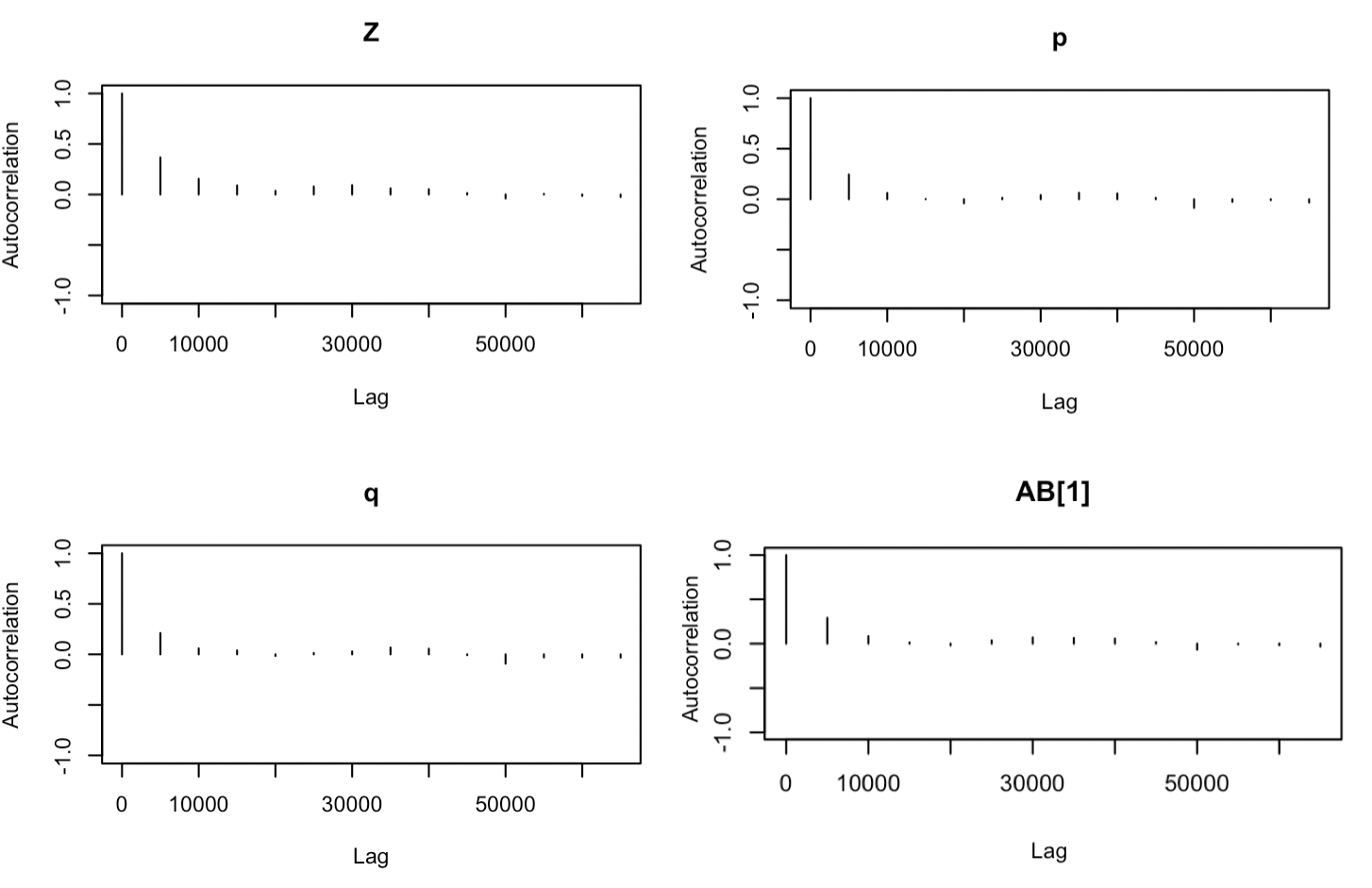}
	\caption[ACF plots of $Z$, $A$, $p$, $q$ from the Bayesian model.] {ACF plots of nodes $Z$, the total number of opioid overdoses, $A$, the number of healthcare-unattended overdoses, branching probabilities $p$, the proportion of healthcare-unattended overdoses, and $q$, the proportion of healthcare-unattended overdose deaths from the Bayesian model on the full opioid tree.}
	\label{fig:opioidACF}
\end{figure}

\clearpage

\bibliography{statrefs}%

\begin{thebibliography}{10}
\providecommand \doibase [0]{http://dx.doi.org/}%

\bibitem{fischeropioid2014}
Fischer B, Keates A, Buhringer G, Remier J, Rehm J. Non-medical use of
  prescription opioid and prescription opioid-related harms: why so markedly
  higher in {North America} compared to the rest of the world?. {\it
  Addiction.} 2014\string; 109:177–81.

\bibitem{guyopioid2017}
Guy J, others . Vital signs: Changes in opioid prescribing in the {United
  States}, 2006-2015. {\it Morb Mortal Wkly Rep.} 2017\string; 66:697–704.

\bibitem{janjuaopioid2018}
Janjua N, others . Identifying injection drug use and estimating population
  size of people who inject drugs using healthcare administrative datasets.
  {\it Int J Drug Policy.} 2018\string; 55:31–39.

\bibitem{overdoses2022}
Canada. oOPHA. Opioid- and Stimulant-related Harms in {Canada}. {\it Federal,
  provincial and territorial Special Advisory Committee on the Epidemic of
  Opioid Overdoses.} 2023\string;
  https://health-infobase.canada.ca/substance-related-harms/opioids-stimulants/.

\bibitem{prescriptionopioidUS}
Chen Q, Larochelle MR, Weaver D, et al. Prevention of Prescription Opioid
  Misuse and Projected Overdose Deaths in the {United States}. {\it JAMA
  Network Open} 2019\string; 2(2):e187621-e187621.
\newblock \href {\doibase :10.1001/jamanetworkopen.2018.7621} {doi:
  :10.1001/jamanetworkopen.2018.7621}

\bibitem{baldwinfent}
Baldwin N, Gray R, Goel A, Wood E, Buxton J, Rieb L. Fentanyl and heroin
  contained in seized illicit drugs and overdose-related deaths in {British
  Columbia}, {Canada}: An observational analysis. {\it Drug and Alcohol
  Dependence} 2018\string; 185:322-327.
\newblock \href {\doibase https://doi.org/10.1016/j.drugalcdep.2017.12.032}
  {doi: https://doi.org/10.1016/j.drugalcdep.2017.12.032}

\bibitem{lifeexp2020}
Canada. S. Table 13-10-0114-01 Life expectancy and other elements of the
  complete life table, three-year estimates, {Canada}, all provinces except
  {Prince Edward Island}. {\it https://doi.org/10.25318/1310011401-eng} 2022.

\bibitem{bccoronerfentanyl2017}
{British Columbia} oCS. Fentanyl-detected illicit drug overdose deaths
  {January} 1, 2012 to {February} 28, 2017. {\it
  http://www.webcitation.org/78lHkI0Bw} 2017.

\bibitem{bccoronerdrug2017}
{British Columbia} oCS. Illicit drug overdose deaths {January} 1, 2007 to
  {March} 31, 2017. {\it http://www.webcitation.org/78lI1v4nV} 2017.

\bibitem{statcan2018}
Canada S. The Daily 2020-01-28 - Life Tables, 2016/2018. {\it
  https://www150.statcan.gc.ca/n1/daily-quotidien/200128/dq200128a-eng.htm}
  2020.

\bibitem{yeopioid2018}
Ye X, Sutherland J, Henry B, Tyndall M, Kendall P. Impact of drug
  overdose-related deaths on life expectancy at birth in {British Columbia}.
  {\it Health Promot Chronic Dis Prev Can.} 2018\string; 38:248–51.

\bibitem{irvinethnkits2018}
Irvine M, Buxton J, Otterstatter M, et al. Distribution of take-home opioid
  antagonist kits during a synthetic opioid epidemic in {British Columbia},
  {Canada}: a modelling study. {\it Lancet Public Health.} 2018\string;
  3:e215-25.

\bibitem{irvinevarbayes2019}
Irvine M, Kuo M, Buxton J, et al. Modelling the combined impact of
  interventions in averting deaths during a synthetic-opioid overdose epidemic.
  {\it Addiction.} 2019\string; 114:1602-1613.

\bibitem{lauraopioiddata}
MacDougall L, Smolina K, Otterstatter M, et al. Development and characteristics
  of the Provincial Overdose Cohort in {British Columbia}, {Canada}. {\it PLoS
  ONE.} 2019\string; 14(1):e0210129.

\bibitem{opioidcanada2019}
Opioid~Overdoses S.~A. C. o. t. E.~oOP. National report: Opioid-related Harms
  in {Canada} Web-based Report. {\it
  https://health-infobase.canada.ca/substance-related-harms/opioids} 2019.

\bibitem{macdougallcohort2019}
Macdougall L, others . Development and characteristics of the Provincial
  Overdose Cohort in {British Columbia}, {Canada}. {\it PLoS One.} 2019\string;
  14(1):e0210129.

\bibitem{flynnmethods}
Flynn M, Gustafson P. Leveraging relational evidence: population size
  estimation on tree-structured data wiht the weighted multiplier method. {\it
  Preprint} 2025\string; Department of Statistics, University of British
  Columbia.

\bibitem{flynncomp}
Flynn M, Gustafson P. {AutoWMM} and {JAGStree} - {R} packages for population
  size estimation on relational tree-structured data. {\it Preprint}
  2025\string; Department of Statistics, University of British Columbia.

\bibitem{bccohortdata}
(2017) BM. {BC Provincial Overdose Cohort}. {V1}. {\it British Columbia
  Ministry of Health [publisher]. Data Extract} 2017\string;
  https://www.popdata.bc.ca/data.

\bibitem{pnetdata}
(2017) B. {PharmaNet}.. {\it British Columbia Ministry of Health [publisher].
  Data Extract} 2017\string; https://www.popdata.bc.ca/data.

\bibitem{nacrsdata}
(2017) C. {National Ambulatory care reporting system (NACRS)}. {\it British
  Columbia Ministry of Health [publisher]. Data Extract} 2017\string;
  https://www.popdata.bc.ca/data.

\bibitem{daddata}
(2017) C. {Discharge Abstract Database}. {\it British Columbia Ministry of
  Health [publisher]. Data Extract} 2017\string;
  https://www.popdata.bc.ca/data.

\bibitem{mspdata}
(2017) BM. {Medical Services Plan (MSP) Payment Information File}. {\it British
  Columbia Ministry of Health [publisher]. Data Extract} 2017\string;
  https://www.popdata.bc.ca/data.

\bibitem{opioiddata}
Graham E, Zhao B, Flynn M, et al. Using Linked Data to Identify Pathways of
  Reporting Overdose Events in {British Columbia}, 2015 - 2017. {\it
  International Journal of Population Data Science} 2022\string; 7:1.

\bibitem{antaIDU2010}
Anta G, Oliva J, Bravo M, De~Mateo S, Domingo-Salvany A. Estimating the
  prevalence of drug injection using a multiplier method based on a register of
  new {HIV} diagnoses. {\it European Journal of Public Health.} 2010\string;
  21(5):646-648.

\bibitem{deangelisIDU2004}
De~Angelis D, Hickman M, Yang S. Estimating Long-Term Trends in the Incidence
  and Prevalence of Opiate Use/Injecting Drug Use and the Number of Former
  Users: Back-Calculation Methods and Opiate Overdose Deaths. {\it American
  Journal of Epidemiology.} 2004\string; 160(10):994-1004.

\bibitem{khalidIDU2014}
Khalid F, Hamad F, Othman A, others . Estimating the number of people who
  inject drugs, female sex workers, and men who have sex with men, {Unguja}
  {Island}, {Zanzibar}: results and synthesis of multiple methods. {\it AIDS
  Behav.} 2014\string; 18:S25-S31.

\bibitem{birrellHIV2013}
Birrell P, Gill O, Delpech V, et al. {HIV} incidence in men who have sex with
  men in {England} and {Wales} 2001-10: a nationwide population study. {\it
  Lancet Public Health.} 2013\string; 13:313-18.

\bibitem{richMSM2017}
Rich A, Lachowsky N, Sereda P, et al. Estimating the size of the MSM population
  in metro {Vancouver}, {Canada}, using multiple methods and diverse data
  sources. {\it J Urban Health.} 2017\string; 95:188-195.

\bibitem{pazbailey2011}
Paz-Bailey G, Jacobson J, Guardado M, others . How many men who have sex with
  mend and female sex workers live in {El Salvador}? Using respondent-driven
  sampling and capture recapture to estimate population sizes. {\it Sex Z
  Transm. Infect.} 2011\string; 87:279-282.

\bibitem{johnstonHIV2011}
Johnston L, Saumtallyb A, Corcealb S, others . High {HIV} and {hepatitis C}
  prevalence amongst injecting drug users in Mauritius: findings from a
  population size estimation and respondent driven sampling survey. {\it Int J
  Drug Policy.} 2011\string; 22:252-258.

\bibitem{heesterbeekreview2015}
Heesterbeek H, Anderson R, Andreasen V, Bansal S, De~Angelis D, others .
  Modeling infectious disease dynamics in the complex landscape of global
  health. {\it Science} 2015\string; 347(6227):aaa4339.

\bibitem{prevostHCV2015}
Prevost T, Presanis A, Taylor A, Golderg D, Hutchinson S, De~Angelis D.
  Estimating the number of people with {hepatitis C} virus who have ever
  injected drugs and have yet to be diagnose: an evidence synthesis approach
  for {Scotland}. {\it Addiction} 2015\string; 110:1287-1300.

\bibitem{sweetingIDU2009}
Sweeting M, De~Angelis D, Ades A, Hickman M. Estimating the prevalence of
  ex-injecting drug use in the population. {\it Stat. Meth. in Med. Research}
  2009\string; 18:381-396.

\bibitem{irvineHIV2018}
Irvine M, Konrad B, Michelow W, Balshaw R, Gilbert M, Coombs D. A novel
  {Bayesian} approach to predicting reductions in {HIV} incidence following
  increased testing interventions among gay, bisexual and other men who have
  sex with men in {Vancouver}, {Canada}. {\it J.R. Soc. Interface} 2018\string;
  15:20170849.

\bibitem{mcdonaldflu2014}
McDonald S, Presanis A, De~Angelis D, et al. An Evidence synthesis appraoch to
  estiamting the incidence of seasonal influenza in the {Netherlands}. {\it
  Influ. Other Respir. Viruses} 2014\string; 8(1):33-41.

\bibitem{karamouzian2019}
Karamouzian M, Kuo M, Crabtree A, Buxton J. Correlates of seeking emergency
  medical help in the even of an overdose in {British Columbia}, {Canada}:
  Findings from the Take Home Naloxone program. {\it Int J Drug Policy.}
  2019\string; 71:157-163.

\bibitem{harmreductionsurvey}
George R, Steinberg A, Buxton J. {BC} Harm Reduction Client Survey: Background
  and Significant Findings throughout the Years. {Vancouver}, {BC}. {BC Centre
  for Disease Control}. {\it
  http://www.bccdc.ca/Health-Professionals-Site/Documents/Harm-Reduction-Reports/}
  2021.

\bibitem{CCENDU}
Drug~Use oCCEN. {CCENDU} Bulletin: Calling 911 in Drug Poisoning Situations.
  {\it
  https://ccsa.ca/sites/default/files/2019-04/CCSA-CCENDU-Calling-911-Drug-Poisoning-2017-en.pdf}
  2017.

\bibitem{towardstheheart}
Moustaqim~Barrette A, Papamihali K, Buxton J. Take Home Naloxone Program
  Report: Review of Data to December 2018. {Vancouver}, {BC}. {BC Centre for
  Disease Control}. {\it https://towardtheheart.com/assets/uploads} 2019.

\bibitem{r}
Team RC. {\it R: A Language and Environment for Statistical Computing}. R
  Foundation for Statistical Computing; Vienna, Austria:  2018.

\bibitem{springergender2012}
Springer K, Mager~Stellman J, Jordan-Young R. Beyond a catalogue of
  differences: A theoretical frame and good practice guidelines for researching
  sex/gender in human health. {\it Soc. Sci. Med.} 2012\string;
  74(11):1817-1824.

\end{thebibliography}


\end{document}